\documentclass[prd,aps,a4,twocolumn,superscriptaddress,preprintnumbers,nofootinbib]{revtex4-1}
\usepackage{pslatex}
\usepackage[pdftex]{graphicx}
\usepackage{psfrag}
\usepackage{epsfig}
\usepackage{color}
\usepackage{cancel}
\usepackage{slashed}
\usepackage{amssymb}
\usepackage{amsmath}
\usepackage{hyperref}
\usepackage{enumerate}
\usepackage{multirow}
\newcommand\subsetsim{\mathrel{\substack{
  \textstyle\subset\\[-0.2ex]\textstyle\sim}}}
\begin{document}
\title{Coherent elastic neutrino-nucleus scattering in multi-ton scale
  dark matter experiments: Classification of vector and scalar
  interactions new physics signals}
%
\author{D. Aristizabal Sierra}%
\email{daristizabal@ulg.ac.be}%
\affiliation{Universidad T\'ecnica
  Federico Santa Mar\'{i}a - Departamento de F\'{i}sica\\
  Casilla 110-V, Avda. Espa\~na 1680, Valpara\'{i}so, Chile}%
\affiliation{IFPA, Dep. AGO, Universit\'e de Li\`ege, Bat B5, Sart
  Tilman B-4000 Li\`ege 1, Belgium}%
\author{Bhaskar Dutta}%
\email{dutta@physics.tamu.edu}%
\affiliation{Department of Physics and Astronomy, Mitchell Institute
  for Fundamental Physics and Astronomy, Texas A\&M University,
  College Station, TX 77843, USA}%
\author{Shu Liao}%
\email{ikaros@physics.tamu.edu}%
\affiliation{Department of Physics and Astronomy, Mitchell Institute
  for Fundamental Physics and Astronomy, Texas A\&M University,
  College Station, TX 77843, USA}%
\author{Louis E. Strigari}%
\email{strigari@physics.tamu.edu}%
\affiliation{Department of Physics and Astronomy, Mitchell Institute
  for Fundamental Physics and Astronomy, Texas A\&M University,
  College Station, TX 77843, USA}%
\begin{abstract}
  We classify new physics signals in coherent elastic neutrino-nucleus
  scattering (CE$\nu$NS) processes induced by $^8$B solar neutrinos in
  multi-ton xenon dark matter (DM) detectors. Our analysis focuses on
  vector and scalar interactions in the effective and light mediator
  limits after considering the constraints emerging from the recent
  COHERENT data and neutrino masses.  In both cases we identify a
  region where measurements of the event spectrum alone suffice to
  establish whether the new physics signal is related with vector or
  scalar couplings. We identify as well a region where 
  measurements of the recoil spectrum are required so to establish the
  nature of the new interaction, and categorize the spectral features
  that enable distinguishing the vector from the scalar case. We
  demonstrate that measurements of the isospin nature of the new
  interaction and thereby removal of isospin related degeneracies are
  possible by combining independent measurements from two different
  detectors. We also comment on the status of searches for vector and
  scalar interactions for on-going multi-ton year xenon experiments.
\end{abstract}

\maketitle
\section{Introduction}
\label{sec:intro}
Next-generation direct detection dark matter (DM) experiments will be
challenged by irreducible solar neutrino backgrounds and eventually,
as exposure increases, by atmospheric neutrino fluxes as well
\cite{Billard:2013qya}. With fairly large portions of the WIMP
parameter space already explored, further exploration of DM direct
detection signals in the near future call for multi-ton size
detectors. Experiments such as XENONnT, LZ and DARWIN
\cite{Aprile:2015uzo,Akerib:2018lyp,Aalbers:2016jon} soon after their
operation will start observing neutrino-induced nuclear recoils, thus
making the identification of an actual signal a difficult task. For
that reason, experimental techniques that enable identifying
background signal events from WIMP-induced recoils have been recently
discussed. They include identification of WIMP signals by their
direction dependencies \cite{Grothaus:2014hja,OHare:2015utx}, time
dependencies induced by WIMP and solar neutrino annual modulation
\cite{Davis:2014ama} and multiple target detectors
\cite{Ruppin:2014bra}.

Although neutrino backgrounds certainly pose a problem for DM
searches, they offer as well various physics opportunities, as they
can be used as a tool for detailed studies of: (i) The coherent
elastic neutrino-nucleus scattering (CE$\nu$NS) process, (ii) solar
and low-energy atmospheric neutrino fluxes, (iii) solar and supernova
physics. Various analyses touching different aspects of these subjects
have been already considered in the literature.  Measurements of the
CE$\nu$NS process in DM detectors will provide complementary
information to that arising from dedicated CE$\nu$NS experiments such
as COHERENT \cite{Akimov:2017ade,Akimov:2018vzs}, CONNIE
\cite{Aguilar-Arevalo:2019jlr}, CONUS \cite{conus} and $\nu$-cleus
\cite{Strauss:2017cuu}. That information can be used to test the
presence of new physics in the form of e.g. neutrino non-standard
interactions (NSI)
\cite{Dutta:2017nht,AristizabalSierra:2017joc,Gonzalez-Garcia:2018dep},
vector or scalar light mediators \cite{Cerdeno:2016sfi} or neutrino
generalized interactions (NGI) \cite{AristizabalSierra:2018eqm}. The
observation of the CE$\nu$NS process will allow a better understanding
of solar and atmospheric neutrino fluxes, the latter poorly understood
with uncertainties of up to order $50\%$. Precise measurements of
low-energy solar neutrino fluxes will in turn improve upon our
understanding of solar physics \cite{Newstead:2018muu}, while a
ton-size detector such as XENONnT will be sensitive to a supernova
burst up to $\sim 35$ kpc from earth, thus providing valuable
information on supernova properties \cite{Lang:2016zhv}.

The CE$\nu$NS and/or the electron-neutrino elastic cross sections are
affected by the presence of new physics in different ways. Since
neutrino-quark NSI are a parametrization of a four-fermion neutral
current process, their effect is just a global rescaling (upwards or
downwards) of the SM differential cross section
\cite{Lindner:2016wff,Dutta:2017nht,AristizabalSierra:2017joc}. Light
mediators interactions (vector or scalar) change that behavior by
introducing an extra momentum transfer dependence, which induces
additional spectral features
\cite{Dent:2016wor,Lindner:2016wff,Farzan:2018gtr,Liao:2017uzy,AristizabalSierra:2019ufd}. Neutrino
electromagnetic couplings can potentially introduce spectral features
as well
\cite{Kosmas:2017tsq,Papoulias:2019txv,Miranda:2019wdy,Miranda:2019skf}. Of
particular interest are neutrino magnetic dipole moments which if
sufficiently large lead to enhancements of the cross section at low
recoil energies. NGI either in the light or effective limits have also
different implications and depending on their nature lead to
distinctive experimental signatures
\cite{Lindner:2016wff,AristizabalSierra:2018eqm,AristizabalSierra:2019ufd,Xu:2019dxe,Bischer:2018zcz,Bischer:2019ttk,Khan:2019jvr}. Given
the number of new physics scenarios and possible signatures that one
could test through measurements of the CE$\nu$NS process, it is
desirable to systematically identify signatures that if observed could
point towards the new physics responsible for a signal.

In this paper we consider such identification in the case of vector
and scalar interactions in the light and effective regimes. For that
aim we consider a xenon-based detector and CE$\nu$NS induced by
$^{8}$B solar neutrinos. We start by writing the interactions we are
interested in Sec. \ref{sec:general} and review the limits to which
they are subject to in Sec. \ref{sec:constraints}. In
Sec. \ref{sec:limits-cevns} we derive as well limits from neutrino
masses that apply on scalar interactions (regardless of the size of
the scalar mediator mass), and that arise through quark condensation
contributing to either the Dirac or Majorana mass operators. Taking
into account these limits, in particular those arising from COHERENT
measurements, we then study the behavior of the signals according to
their parameter space dependence in Sec. \ref{sec:vec-scalar}. We
first identify cases in which vector and scalar interactions can be
distinguished by measurements of the event spectrum alone. We then
identify cases in which combined measurements of the event and recoil
spectrum are required. We evaluate as well the capability of multi-ton
scale DM detectors to determine the isospin nature of the new physics
signal (using silicon, argon and germanium in addition to xenon) in
Sec. \ref{sec:isospin-cons-viol}. We pay special attention to the case
of degeneracies in xenon and determine the most suited nuclide for
parameter degeneracy breaking in Sec. \ref{sec:degeneracies}. Finally
in Sec. \ref{sec:conclusions} we present our conclusions.

\begin{table*}
  \centering
  \renewcommand{\arraystretch}{1.3}
  \begin{tabular}{|c|c|c||c|c||c|c|c|c||c|c|c|c|c|c|}\hline
    &\multicolumn{2}{|c||}{Silicon}
    &\multicolumn{2}{|c||}{Argon}
    &\multicolumn{4}{|c||}{Germanium}
    &\multicolumn{6}{|c|}{Xenon}\\\hline\hline
    \multirow{3}{*}{Nuc}
    &{\footnotesize$^{28}$Si}&{\footnotesize$9.22\cdot 10^{-1}$}
    &{\footnotesize$^{36}$Ar}&{\footnotesize$3.37\cdot 10^{-3}$}
    &{\footnotesize$^{70}$Ge}&{\footnotesize$2.04\cdot 10^{-1}$}
    &{\footnotesize$^{72}$Ge}&{\footnotesize$2.73\cdot 10^{-1}$}
    &{\footnotesize$^{124}$Xe}&{\footnotesize$9.50\cdot 10^{-4}$}
    &{\footnotesize$^{126}$Xe}&{\footnotesize$8.90\cdot 10^{-4}$}
    &{\footnotesize$^{128}$Xe}&{\footnotesize$1.91\cdot 10^{-2}$}\\\cline{2-15}
    &{\footnotesize$^{29}$Si}&{\footnotesize$4.68\cdot 10^{-2}$}
    &{\footnotesize$^{38}$Ar}&{\footnotesize$6.32\cdot 10^{-4}$}
    &{\footnotesize$^{73}$Ge}&{\footnotesize$7.76\cdot 10^{-2}$}
    &{\footnotesize$^{74}$Ge}&{\footnotesize$3.67\cdot 10^{-1}$}
    &{\footnotesize$^{129}$Xe}&{\footnotesize$2.64\cdot 10^{-1}$}
    &{\footnotesize$^{130}$Xe}&{\footnotesize$4.07\cdot 10^{-2}$}
    &{\footnotesize$^{131}$Xe}&{\footnotesize$2.12\cdot 10^{-1}$}\\\cline{2-15}
    &{\footnotesize$^{30}$Si}&{\footnotesize$3.09\cdot 10^{-2}$}
    &{\footnotesize$^{40}$Ar}&{\footnotesize$9.96\cdot 10^{-1}$}
    &{\footnotesize$^{76}$Ge}&{\footnotesize$7.83\cdot 10^{-2}$}
    &{\footnotesize---}&{\footnotesize---}
    &{\footnotesize$^{132}$Xe}&{\footnotesize$2.69\cdot 10^{-1}$}
    &{\footnotesize$^{134}$Xe}&{\footnotesize$1.04\cdot 10^{-1}$}
    &{\footnotesize$^{136}$Xe}&{\footnotesize$8.86\cdot 10^{-2}$}\\\cline{1-15}
    $m_N$ [GeV/c$^2$]
    &\multicolumn{2}{|c||}{$26.16$}
    &\multicolumn{2}{|c||}{$37.21$}
    &\multicolumn{4}{|c||}{$67.66$}
    &\multicolumn{6}{|c|}{$122.29$}\\\hline
    $A_N$
    &\multicolumn{2}{|c||}{$28.10$}
    &\multicolumn{2}{|c||}{$39.98$}
    &\multicolumn{4}{|c||}{$72.70$}
    &\multicolumn{6}{|c|}{$131.39$}\\\hline
  \end{tabular}
  \caption{Silicon, argon, germanium and xenon stable isotopes 
    along with their relative abundances. Nuclear mass and mass 
    numbers are calculated by averaging over the relative 
    abundance of each isotope.}
  \label{tab:nuclei-parameters}
\end{table*}
\section{Vector and scalar neutrino 
  generalized interactions}
\label{sec:general}
Vector and scalar NGI scenarios are dictated by the following
interactions \cite{AristizabalSierra:2019ufd}
\begin{equation}
  \label{eq:vector-NGI-light}
  \mathcal{L}^V=\overline{\nu}\gamma_\mu(f_V+if_A\gamma_5)\nu\,V^\mu
  + \sum_{q=u,d}h_V^q\overline{q}\gamma_\mu q\,V^\mu
\end{equation}
and
\begin{align}
  \label{eq:scalar-NGI-light}
  \mathcal{L}^S_\text{LNC}&=\overline{\nu}(f_S+if_P\gamma_5)\nu\,S
  + \sum_{q=u,d}h_S^q\overline{q}q\,S\ ,
  \nonumber\\
  \mathcal{L}^S_\text{LNV}&=\overline{\nu^c}(f_S+if_P\gamma_5)\nu\,S
  + \sum_{q=u,d}h_S^q\overline{q}q\,S\ ,  
\end{align}
where in the scalar case lepton number conserving (LNC) couplings
require the presence of right handed neutrinos.  Note that the
couplings can involve CP violating phases that we do not consider (see
Ref. \cite{AristizabalSierra:2019ufd} for an analysis including CP
violation). The quark sector can involve as well axial and
pseudoscalar currents. These couplings lead to nuclear spin-dependent
processes which are suppressed compared with those induced by vector
and scalar quark currents. The CE$\nu$NS cross sections induced by the
interactions in (\ref{eq:vector-NGI-light}) and
(\ref{eq:scalar-NGI-light}) are given by
\begin{align}
  \label{eq:x-sec-V}
  \frac{d\sigma_V}{dE_r}=&\frac{G_F^2}{2\pi}m_N
  |\xi_V|^2\left(2 - \frac{E_rm_N}{E_\nu^2}\right)
  F^2(q^2)\ ,
  \\
  \label{eq:x-sec-S}
   \frac{d\sigma_S}{dE_r}=&\frac{G_F^2}{2\pi}m_N
   |\xi_S|^2\frac{E_rm_N}{2E_\nu^2}\,F^2(q^2)\ ,
\end{align}
where $E_r$ refers to nuclear recoil energy
($E_r^\text{max}\simeq 2E_\nu^2/m_N$, with $E_\nu$ the ingoing
neutrino energy). Note that here we have assumed the same nuclear form
factor for protons and neutrons. Such a choice is accurate provided
one assumes the root-mean-square (rms) radii of the neutron and proton
distributions are equal. Possible deviations from this
assumption---allowed by uncertainties on the rms radius of the neutron
distribution---require the proton and neutron contributions to be
weighted by their own from factors
\cite{AristizabalSierra:2019zmy}. For our analysis we use the Helm
form factor \cite{Helm:1956zz}. For the rms radii of the proton
distributions of xenon, silicon and germanium we use the average
$\langle r_k\rangle=\sum_k X_k r_k$, where $r_k$ refers to the rms
radius of the $k$-th isotope \cite{Angeli:2013epw} \footnote{Choosing
  a different value will not sizably affect our results. For $^8$B
  neutrino energies, the form factor approaches 1.}. The new physics
couplings are encoded in $\xi_V$ and $\xi_S$ which read
\begin{align}
  \label{eq:xiV-xiS}
  \xi_V&=g_V + \frac{C_V^NF_V}{\sqrt{2}G_F(2m_NE_r+m_V^2)}\ ,
  \nonumber\\
  \xi_S&=\frac{C_S^NF_S}{G_F(2m_NE_r+m_S^2)}\ ,
\end{align}
with $F_V=f_V-if_A$, $F_S=f_S-if_P$. The neutrino-nucleus vector
couplings $C_V^N$ and $g_V$ (SM contribution) as well as the
neutrino-nucleus scalar parameter are obtained by going from the quark
to the nucleus operators. They are written as
\cite{AristizabalSierra:2018eqm}
\begin{align}
  \label{eq:fundamental-couplings-nuclear-couplings-vector}
  C_V^N&=Z(2h_V^u + h_V^d) + (A-Z)(h_V^u + 2h_V^d)\ ,
  \\
  g_V&=Z(2g_V^u + g_V^d) + (A-Z)(g_V^u + 2g_V^d)\ ,
  \\
  \label{eq:fundamental-couplings-nuclear-couplings-scalar}
  C_S^N&=Z\sum_{q=u,d}h_S^q\frac{m_p}{m_q}f_{T_q}^p
       +
       (A-Z)\sum_{q=u,d}h_S^q\frac{m_n}{m_q}f_{T_q}^n\ .
\end{align}
Here $g_V^u=1/2-4/3\sin^2\theta_W$, $g_V^d=-1/2+2/3\sin^2\theta_W$ and
$\sin^2\theta_W=0.231$ \cite{1674-1137-40-10-100001}. In
(\ref{eq:fundamental-couplings-nuclear-couplings-scalar})
contributions from the strange and heavy quarks have been neglected.
Values for the hadronic form factors $f_{T_q}^{p,n}$ are derived in
chiral perturbation theory from measurements of the $\pi$-nucleon
sigma term
\cite{Crivellin:2013ipa,Hoferichter:2015dsa,Ellis:2018dmb}. Updated
values are given by \cite{Hoferichter:2015dsa}
\begin{alignat}{2}
  \label{eq:ftqn}
  f_{T_u}^p&=(20.8\pm 1.5)\times 10^{-3}\ ,\;&
  f_{T_d}^p=(41.1\pm 2.8)\times 10^{-3}\ ,
  \nonumber\\
  f_{T_u}^n&=(18.9\pm 1.4)\times 10^{-3}\ ,\;&
  f_{T_d}^n=(45.1\pm 2.7)\times 10^{-3}\ .
\end{alignat}
Note that the new physics couplings in (\ref{eq:xiV-xiS}) reduce to
effective couplings for $m_X^2\gg 4E_\nu^2$ ($X=V,S$), which for
$E_\nu\lesssim 100\,$MeV---as required by coherence of the
neutrino-nucleus elastic scattering process---means that for
$m_X\gtrsim 10^3\,$MeV CE$\nu$NS induced by the interactions
in~(\ref{eq:vector-NGI-light}) and (\ref{eq:scalar-NGI-light}) is well
described by the four-point contact interactions studied in
Refs. \cite{Lindner:2016wff,AristizabalSierra:2018eqm,Bischer:2019ttk}
(effective NGI) with
\begin{equation}
  \label{eq:eff-coupl}
  \xi_V^\text{BSM}\to \frac{C_V^NF_V}{\sqrt{2}G_F m_V^2}\ ,\qquad
  \xi_S\to \frac{C_S^NF_S}{G_F m_S^2}\ .
\end{equation}
More precise numbers are given in the Sec. \ref{sec:limits-cevns}.
\subsection{Recoil spectrum and event rate}
\label{sec:recoil-spectrum-event-rate}
We will consider silicon, argon, germanium and xenon detectors. For
silicon, germanium and xenon the event rate spectrum comprises
contributions from all their stable isotopes (see
Tab. \ref{tab:nuclei-parameters}). For argon only the effects of
$^{40}$Ar are relevant, given that its relative abundance amounts to
$99.6\%$. In our analysis rather than using the contributions from all
isotopes we assume a single contribution by fixing the nuclear mass
and mass number according to $\langle m_N\rangle=\sum_i X_i m_i$ and
$\langle A\rangle=\sum_i X_i A_i$. Here $m_i$ refers to the mass of
the $i$-th isotope in GeV/c$^2$, $A_i$ to its mass number and $X_i$ to
its relative abundance.

The recoil spectrum can then be written as
\begin{equation}
  \label{eq:event-rate}
  \frac{dR}{dE_r}=\frac{N_A}{\langle A\rangle}
  \int_{E_\nu^{min}}^{E_\nu^{max}}\Phi(E_\nu)\frac{d\sigma}{dE_r} dE_\nu\ ,
\end{equation}
where $N_A=6.022\times 10^{23}\;\text{kg}^{-1}$ and $\Phi(E_\nu)$ the
neutrino flux. The lower integration limit is given by
$E_\nu^\text{min}=\sqrt{\langle m_N\rangle E_r/2}$, while the upper
limit by the kinematic endpoint of the corresponding neutrino
spectrum. The total number of events follows from integration of the
recoil spectrum
\begin{equation}
  \label{eq:events}
  N_\text{events}=\int_{E_r^\text{min}}^{E_r^\text{max}}\frac{dR}{dE_r}
  \mathcal{A}(E_r)\,dE_r\ ,
\end{equation}
where $\mathcal{A}(E_r)$ refers to the experimental acceptance. For a
binned analysis limits of integration are determined by bin width
according to $E_r\pm \Delta E_r$.
\begin{figure*}
  \centering
  \includegraphics[scale=0.37]{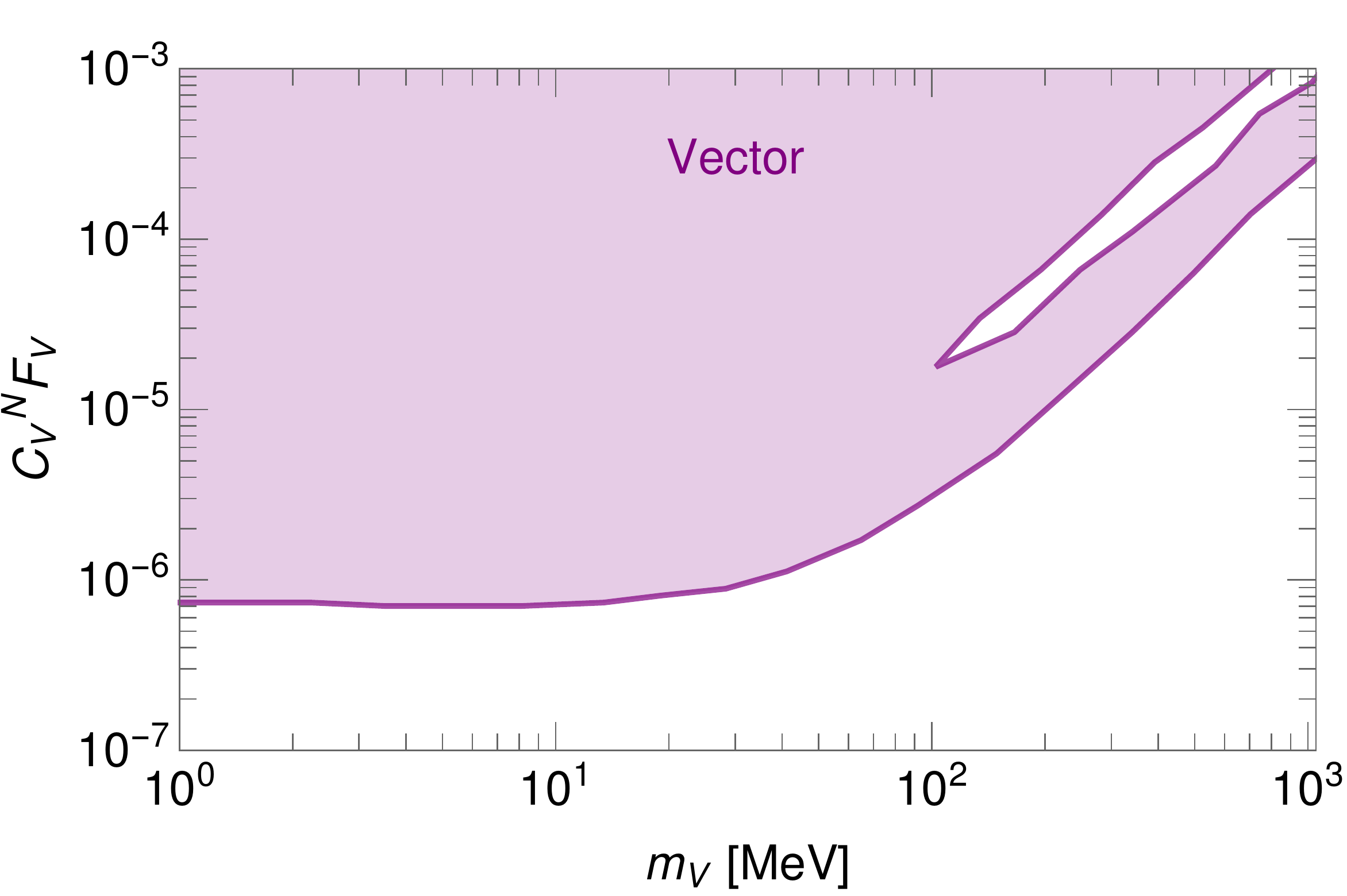}
  \hfill
  \includegraphics[scale=0.37]{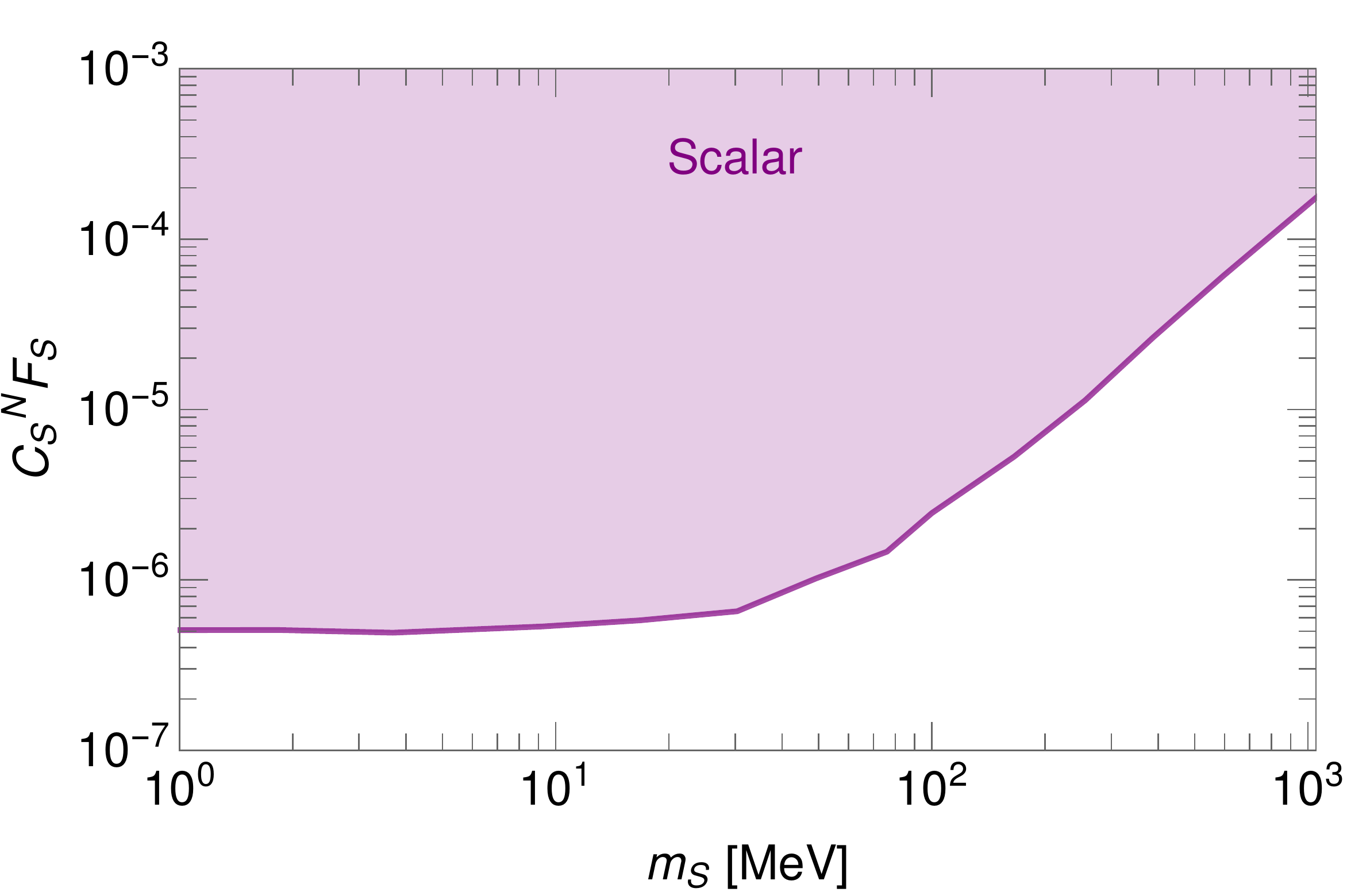}
  \caption{\textbf{Left graph}: 90\% CL limits on vector couplings for
    light vector mediator scenarios from COHERENT data assuming an
    energy flat quenching factor \cite{Akimov:2017ade}, using spectral
    and time information and derived through a likelihood
    analysis. \textbf{Right graph}: Same as left graph but for
    scalars.}
  \label{fig:coh-limits}
\end{figure*}
\subsection{Constraints on vector and scalar couplings}
\label{sec:constraints}
In this section we first discuss laboratory limits and then
constraints from COHERENT.  We then discuss bounds for light mediator
scenarios arising from astrophysical and cosmological observations,
most of them subject to fairly large uncertainties. Particularly
useful for our analysis are the bounds arising from COHERENT.
\subsubsection{Laboratory limits}
\label{sec:limits-lab}
Laboratory limits include bounds from fixed target and beam dump
experiments, rare charged lepton decays, accelerator and neutrino
data. They have been recently analyzed in
Ref. \cite{Bauer:2018onh}. Here we briefly summarize them and discuss
why for interactions (\ref{eq:vector-NGI-light}) and
(\ref{eq:scalar-NGI-light}) they can easily be evaded. Apart from
accelerator searches, these limits apply only in the case of light
mediators due kinematic constraints. Light vector or scalar bosons, if
light enough, can be produced in the collision of protons with fixed
targets.  Production of $V$ can proceed via Bremsstrahlung or via pion
production and subsequent decay, $\pi^0\to \gamma + V$, while
production of $S$ only through Bremsstrahlung. Limits derived from
these types of experiments rely on charged lepton decay modes, which
the interactions in (\ref{eq:vector-NGI-light}) and
(\ref{eq:scalar-NGI-light}) generate only at the one-loop order. These
limits therefore can be safely ignored.

Limits from rare charged lepton decays proceed from muon and tau
decays modes comprising $V$ or $S$ in the final state, such as
$\mu^+\to e^+ \nu_e \bar\nu_\mu + X$ ($X=S,V$). As in the case of
fixed target and beam dump experiments, limits derived from these type
of processes require the new state to decay to charged leptons and so
in our case these bounds can be readily evaded. LHC limits arise from
Drell-Yan production, Higgs and $D$ meson decays (ATLAS, CMS and LHCb
\cite{Aaij:2017rft}). At $e^+e^-$ colliders (KLOE, BaBar and Belle-II
\cite{Anastasi:2016ktq,Lees:2014xha,Inguglia:2016acz}) through
radiative return and heavy meson decays. Relevant to our case is only
Drell-Yan production which applies for vector or scalar masses above
$12\;$GeV. Detection is done by looking for opposite-charge lepton
pairs \cite{Aaij:2017rft,Curtin:2014cca}, and so again these bounds we
can safely ignore. Same arguments apply for bounds derived from
neutrino data, which include neutrino trident production
\cite{Dorenbosch:1986tb,Mishra:1991bv,Zeller:2001hh}, Borexino
\cite{Kaneta:2016uyt} and Texono
\cite{Bilmis:2015lja,Lindner:2018kjo}, and which require couplings of
the vector or scalar to charged leptons. In summary, in some cases the
interactions in Eqs. (\ref{eq:vector-NGI-light}) and
(\ref{eq:scalar-NGI-light}) allow the production of the vector or
scalar bosons. However since detection relies on charged lepton decay
modes, laboratory constraints in our case are loop suppressed.

For heavy states, bounds from muon and tau lepton flavor-violating
decay modes ($\mu-e$ conversion in nuclei, $\mu\to e\gamma$ and
$\tau\to \rho \ell$, $\ell=e,\mu$), contact interactions and violation
of universality could severely constraint the available parameter
space \cite{Wise:2014oea,Bischer:2019ttk}. This however requires
couplings to charged leptons and so these bounds in our case can be
evaded too.
\subsubsection{Limits from neutrino scattering: CE$\nu$NS}
\label{sec:limits-cevns}
In general neutrino scattering data can be used to search for new
physics or otherwise to set limits on new interactions. Sensitivities
depend---of course---on the quality of the available data as well as
on the uncertainties that the neutrino-$\mathcal{N}$ cross section
involves ($\mathcal{N}$ stands for nucleus or nucleon, depending on
the incoming neutrino energy). The CE$\nu$NS energy domain is
determined by the coherence condition $q\lesssim R_N^{-1}$, and
depending on the target material it is roughly below 100 MeV. The
quality of the COHERENT data combined with a cross section with
relatively small nuclear uncertainties \footnote{Mainly dominated by
  the lack of experimental information on the root-mean-square radius
  of the neutron distribution \cite{AristizabalSierra:2019zmy}.},
makes CE$\nu$NS a rather powerful tool.

\begin{figure*}
  \centering
  \includegraphics[scale=1.]{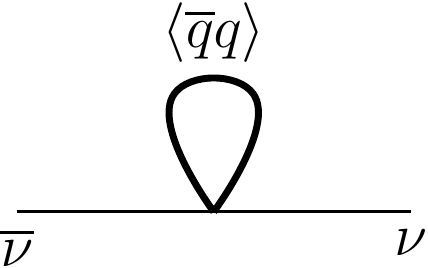}
  \caption{Contribution to neutrino masses from quark condensates
    induced by scalar interactions. The contribution can be of Dirac
    or Majorana type depending on whether the scalar coupling is or
    not lepton number violating. The loop refers to quark
    condensates. It is just a diagrammatic representation and so no
    loop suppression factor is involved.}
  \label{fig:nmm-contribution}
\end{figure*}
Depending on the energy window and on the neutrino energy and mediator
mass relative size, other neutrino scattering processes can play a
rather important role for $E_\nu\gtrsim 0.1\;$GeV. For
$E_\nu \simeq 0.1-20\;\text{GeV}$ a number of scattering processes are
relevant. They include neutrino quasi-elastic scattering, neutral
current elastic scattering, resonant single pion production and
coherent pion production. Recent measurements of these processes
involve data from MiniBooNE
\cite{AguilarArevalo:2010zc,Aguilar-Arevalo:2013dva,AguilarArevalo:2010cx},
NOMAD \cite{Kullenberg:2009pu} and MINER$\nu$A
\cite{Fiorentini:2013ezn}, among others. Constraints on new physics
from these processes however are nonexistent. The reason could be
related with the fact that these cross sections are subject to
relatively large nuclear effects uncertainties.

For $E_\nu\simeq 20-500\;$GeV, the domain of deep inelastic scattering
(DIS), the neutrino interacts with a quark in the nucleon. Data comes
from CHARM-II \cite{Dorenbosch:1986tb} and NuTeV \cite{Zeller:2001hh}
and the cross section is subject to relatively small uncertainties,
compared to the previous processes. Bounds for vector interactions
(NSI) in the limit $m_V^2\gg 2 E_\nu m_N x y$ ($x$ refers to the
Bjorken variable, $y$ to the inelasticity parameter and $m_N$ to the
nucleon mass), have been derived in \cite{Coloma:2017egw}. DIS limits
are nonexistent for scalar interactions nor for vector interactions in
the mass range $m_V^2\simeq 2 E_\nu m_N x y$.

Focusing then on CE$\nu$NS, constraints on scalar and vector
interactions in the light and effective regimes have been derived in a
series of papers using COHERENT data
\cite{Liao:2017uzy,Kosmas:2017tsq,AristizabalSierra:2018eqm,Farzan:2018gtr}.
Recently using the Chicago-3 quenching factor \cite{Collar:2019ihs}
Refs. \cite{Papoulias:2019txv,Khan:2019cvi} updated those limits. We,
however, recalculated them keeping the original quenching factor, but
follow an analyses which includes not only spectral information but
temporal information as well~\cite{Dutta:2019eml,Giunti:2019xpr}. And
rather than adopting a chi-square test implement a likelihood
statistical analysis (see below). The results are displayed in
Fig. \ref{fig:coh-limits}, which show the 90\% CL limits in the
$C_X^NF_X-m_X$ plane. In there one can see that at
$m_X\gtrsim 30\;$MeV the effective limit starts kicking in and at
$m_X\simeq 10^3\;$MeV CE$\nu$NS is already dominated by it.

To derive these results we have used the spectral neutrino functions
\begin{align}
  \label{eq:spectral-nu-fun}
  \mathcal{F}_{\nu_\mu}(E_\nu)&=\frac{2m_\pi}{m_\pi^2-m_\mu^2}
                               \delta\left(1 - \frac{2E_\nu}{m_\pi^2-m_\mu^2}\right)\ ,
                               \nonumber\\
  \mathcal{F}_{\nu_e}(E_\nu)&=\frac{192}{m_\mu}\left(\frac{E_\nu}{m_\nu}\right)^2
                               \left(\frac{1}{2} - \frac{E_\nu}{m_\mu}\right)\ ,
                               \nonumber\\
  \mathcal{F}_{\bar\nu_\mu}(E_\nu)&=\frac{64}{m_\mu}\left(\frac{E_\nu}{m_\nu}\right)^2
                               \left(\frac{3}{4} - \frac{E_\nu}{m_\mu}\right)\ ,
\end{align}
which are then normalized to $\mathcal{N}=r\,n_\text{POT}/(4\pi L^2)$,
with $r=0.08$, $n_\text{POT}=1.76\times 10^{23}$ and $L=19.3\;$m.
Recoil energy binning is determined by number of photoelectrons
$n_\text{PE}$ which are related with $E_r$ through
$n_\text{PE}=1.17(E_r/\text{keV})$
\cite{Akimov:2017ade,Akimov:2018vzs}. For the data analysis we define
our likelihood function as:
\begin{align}
  \label{eq:L}
  \mathcal{L}(\vec\theta|t,E_r)&=
                                 \mathfrak{N}
                                 \prod_{(t,E_r)}\int d\alpha \sum_{N_\text{bg}}
                                 P(N_\text{obs},\lambda)P(N_\text{obs,bg},N_\text{bg})
                                 \nonumber\\
  &\quad\times G(\alpha,\sigma_\alpha^2)\ ,
\end{align}
where $P(n,\nu)=\nu^ne^{-\nu}/n!$ refers to a Poisson distribution
function, while $G(x,\sigma^2)=e^{-x^2/2\sigma^2}/\sqrt{2\pi\sigma^2}$
to a Gaussian distribution function with zero mean. The new physics
parameters are encoded in the parameter space vector $\vec \theta$,
$\mathfrak{N}$ is a normalization factor that assures unit
normalization of the likelihood function when integrated over
$\vec\theta$,
$\lambda(t,E_r)=(1+\alpha)N(t,E_r,\vec\theta) + N_\text{bg}(t,E_r)$,
is the expected/observed number of events with uncertainty parameter
$\alpha$ accounting for the systematic uncertainties from flux,
nuclear form factor, quenching factor and signal acceptance. From our
notation in (\ref{eq:L}) it is clear that we assume this parameter
follows a Gaussian distribution with zero mean and standard deviation
$\sigma_\alpha=0.28$ \cite{Akimov:2017ade}.  $N(t,E_r,\vec\theta)$ is
the number of neutrino-induced recoil events predicted by theory and
derived from (\ref{eq:events}) using Eqs. (\ref{eq:x-sec-V}) and
(\ref{eq:x-sec-S}). $N_\text{bg}(t,E_r)$ is the true background count
(not observed by definition), while $N_\text{obs,bg}(t,E_r)$ is the
observed background reported by the COHERENT experiment. We therefore
integrate over $N_\text{bg}(t,E_r)$ assuming a flat prior
distribution. We consider both energy and timing spectra by binning
the data with 2 photoelectrons in recoil energy space and $0.5\,\mu s$
in time space. Further details of this analysis can be found for the
vector case in Ref. \cite{Dutta:2019eml}.

Maximal enhancement of vector and scalar interactions happen for
$m_X=1\;$MeV. For that value the exclusion plots in
Figs. \ref{fig:coh-limits} fix the nuclear and neutrino couplings to
$C_V^NF_V|_\text{Exp}\leq 7.4\times 10^{-7}$ and
$C_S^NF_S|_\text{Exp}\leq 5.1\times 10^{-7}$. The value of the nuclear
coupling depends on the nuclear target. For $^{133}$Cs, for which the
fit has been done, one has
\begin{align}
  \label{eq:nuclear-couplings-cs}
  C_V^NF_V&=\left(211 h_V^d + 188 h_V^u\right)F_V ,
            \nonumber\\
  C_S^NF_S&= \left(1154.54 h_S^d + 1117.53 h_S^u\right)F_S\ .
\end{align}
Assuming $h_X^q=h_X$, i.e. assuming isospin conserving (violating)
vector (scalar) interactions, bounds on the fundamental couplings read
$h_V\times F_V\leq 2.0\times 10^{-9}$ and
$h_S\times F_S\leq 2.2\times 10^{-10}$. Larger values are possible if a
certain degree of fine tuning is at work, but we will not consider
such possibility. With these numbers it becomes clear that the lighter
(heavier) the isotope the less (more) prominent the effects, with the
suppression (enhancement) given by $A/A_\text{Cs}$.

We calculate the limits in the effective couplings at 90\% CL using
COHERENT data with both temporal and energy spectrum:
\begin{align}
  \label{eq:limits-effective}
  \widetilde{\xi}_V&=[1.06,13.01]\oplus [64.81,74.77]\ ,
  \nonumber\\
  \widetilde{\xi}_S&=[-16.08,16.08]\ .
\end{align}
Notice that the 90\%CL for effective couplings does not contains SM
($\widetilde{\xi}_V=0$), this is because including temporal
information of the COHERENT data gives signal of non-standard
interaction at about $2\sigma$ level as discovered in
Ref. \cite{Dutta:2019eml}.
\subsubsection{Limits on scalar interactions from neutrino masses}
\label{sec:nu-mass-limits}
Scalar interactions are subject as well to constraints from neutrino
masses. Below $\Lambda_\text{QCD}\simeq 200\,$MeV quark condensates
$\langle\bar qq\rangle$ induce a contribution to neutrino masses as
depicted in Fig.~\ref{fig:nmm-contribution}. Since the neutrino mass
operator is calculated at $m_\nu=\Sigma(p^2=0)$, the contribution is
relevant regardless of the scalar mass.  The overall mass scale of the
neutrino mass matrix is determined by $G_F\langle\bar qq\rangle$,
whereas the lepton flavor structure by the couplings $F_S$. The quark
condensate can be evaluated in the Nambu-Jona-Lasinio model which
gives $\langle\overline{q}q\rangle=(8\pi/\sqrt{3})f_\pi$
\cite{Nambu:1961tp}, with the value of the pion decay constant given
by $f_\pi=89.8\,$MeV as measured from the charged pion decay lifetime
\cite{Bernstein:2011bx}. The neutrino mass contribution from the
diagram in Fig.~\ref{fig:nmm-contribution} can thus be written as
\begin{equation}
  \label{eq:nmm-cont}
  m_\nu=\frac{8\pi}{\sqrt{3}}G_Ff_\pi\,F_S\sum_qh_S^q
  \simeq 122.5\,\text{eV}\;F_S\sum_qh_S^q\ .
\end{equation}
At the 95\% CL cosmological limits on neutrino masses vary from
$\sum m_\nu<0.6\,$eV to $\sum m_\nu<0.12\,$eV depending on the data
sets used. The most stringent bound is obtained by including baryon
acoustic oscillation data \cite{Aghanim:2018eyx}. This limit combined
with (\ref{eq:nmm-cont}) implies
\begin{equation}
  \label{eq:neutrino-mass-limit}
  F_S\times \sum_qh_S^q<9.79\times 10^{-4}
  \quad \text{(neutrino mass limit)}\ .
\end{equation}
It can be satisfied with large $h_S^q$ and suppressed $F_S$ or vice
versa. Large values for $F_S$ and $h_S^q$ are possible too, but a
delicate cancellation between $h_S^u$ and $h_S^d$ is required. As
discussed in the previous section, for $m_S\lesssim 10^3\,$MeV
COHERENT constraints are more competitive. Indeed given the values
that the couplings can have in that mass window, scalar interactions
through quark condensates cannot sizably contribute to $\sum
m_\nu$.
Thus, if a signal of this type of interactions is observed and one can
establish $m_S\lesssim 10^3\,$MeV, one can be sure that neutrino mass
generation should proceed through a different mechanism.

For $m_S\gtrsim 10^3\,$MeV one is already in the effective limit where
scalar interactions are controlled by the scalar parameter
in~(\ref{eq:eff-coupl}). In terms of the quark couplings this
parameter is maximized in the limit $h_S^u\to h_S^d\to h_S$ for which,
using the central values of the hadron form factors in
(\ref{eq:ftqn}), $C_S^N$ can be written as
\begin{equation}
  \label{eq:CSN}
  C_S^N=h_S\,\left[16.54(A-Z) - 16.58 Z\right]\ .
\end{equation}
Thus, combined with the neutrino mass limit the strength of the scalar
interaction compared with the SM contribution is bounded as follows
\begin{align}
  \label{eq:limits-nmasses}
  \text{Xe}:&\,\, \xi_S\lesssim \left(\frac{302.2}{m_S/\text{GeV}}\right)^2\ ,\quad
  \text{Ge}:\,\,\xi_S\lesssim \left(\frac{224.8}{m_S/\text{GeV}}\right)^2\ ,
             \nonumber\\
  \text{Ar}:&\,\,\xi_S\lesssim \left(\frac{166.6}{m_S/\text{GeV}}\right)^2\ ,\quad
  \text{Si}:\,\,\xi_S\lesssim \left(\frac{139.8}{m_S/\text{GeV}}\right)^2\ .
\end{align}
As expected from (\ref{eq:CSN}) the limit is isotope dependent, but
the differences between heavy (xenon), intermediate (germanium) and
light (silicon and argon) nuclides is at most a factor $\sim 2$. The
result in (\ref{eq:limits-nmasses}) demonstrates that only for
$m_S\gtrsim 300\,$GeV neutrino mass limits lead to suppressed scalar
couplings. Indeed, they show that if one takes into account only that
constraint the scalar interaction can be way larger than the SM
contribution when $m_S\lesssim 300\,$GeV.
\subsubsection{Astrophysical limits}
\label{sec:add-limits}
Scalar interactions are subject to further constraints that apply on
either quark or neutrino couplings or both simultaneously. They were
recently discussed in Ref. \cite{Farzan:2018gtr} and below we
summarize them. These limits can be sorted in three different groups
depending on how they affect neutrino properties within the supernova
(SN) inner core. Since neutrinos are trapped in the SN core they can
only escape by diffusion, with the diffusion time determined by
$t_\text{diff}\simeq R_\text{SN}^2/\lambda$ ($R_\text{SN}$ is the SN
core radius and $\lambda$ the neutrino mean free path). Regardless of
whether the interaction is or not lepton number violating its presence
can reduce $t_\text{diff}$ by increasing $\lambda$. Assuring that
neutrino trapping is not strongly disrupted, i.e. that $t_\text{diff}$
does not decreases below $\sim 10$\,secs, translates into an upper
bound on scalar couplings that depends on the scalar mass,
$C_S^NF_S\lesssim 1.2\times 10^{-7}$ for $m_S=1\;$MeV and about an
order of magnitude larger for $m_S=100\;$MeV.

In the LNC case, more stringent bounds follow from SN cooling and
sterile neutrino trapping, with the strongest bound arising from the
former. Since sterile neutrinos are not subject to electroweak
interactions, once they are produced they can escape the SN core, thus
leading to fast energy loss if the active-sterile neutrino rate
conversion is high enough. In the LNV case, instead, the new
interaction can modify the electron neutrino chemical potential for
sufficiently large couplings, thus affecting the SN equation of state;
something that can be understood fully in terms of the electron
neutrino chemical potential. In the absence of new interactions an
electron neutrino asymmetry is present ($\mu_{\nu_e}\neq 0$, with
$\mu_{\nu_e}$ the electron neutrino chemical potential). If the LNV
interaction attains thermal equilibrium it will enforce
$4\mu_{\nu_e}=\mu_S$, which implies $n_{\nu_e}=n_{\bar\nu_e}$ given
that $\mu_S=0$ \footnote{The generation of an asymmetry in $S$ (a
  chemical potential) requires $S\neq S^*$, departure from thermal
  equilibrium and a CP-violating interaction.}. In other words, if the
new couplings are large enough the interaction will tend to
equilibrate $n_{\nu_e}$ and $n_{\bar\nu_e}$, affecting the SN equation
of state. The limits derived from these arguments depend on the scalar
mass. For $m_S=1\;$MeV in the LNC case the most stringent limit reads
$C_S^NF_S\lesssim 3.3\times 10^{-9}$, while for LNV interactions
$C_S^NF_S\lesssim 2.6\times 10^{-10}$.

Light vectors are subject to limits from stellar cooling and SN
arguments as well.  Since the temperature of the helium core in
horizontal branch stars is of order
$10^8\;\text{K}\simeq 10^{-2}\;\text{MeV}$, vector bosons with masses
up to $0.1\;$MeV can be produced through $^{4}$He Compton scattering
processes (this value possible from the energy tail of the
distribution). Avoiding energy loss through these processes implies
$h_V^{p,n}\lesssim 4\times 10^{-11}$
\cite{Grifols:1986fc,Grifols:1988fv}. Constraints from disruption of
the neutrino diffusion time in SN apply as well \cite{Chang:2016ntp}.

There is however few caveats on these bounds that one should bear in
mind. First of all, uncertainties on core-collapse SN are still
large. Limits derived from SN arguments therefore should be understood
as order of magnitude estimations \cite{Muller:2016izw}. For vectors,
stellar cooling arguments ignore plasma mixing effects, considering
them results in different bounds \cite{Hardy:2016kme}. And finally,
these limits can be avoided if the new states couple to scalars that
condensate inside the star or the SN core. In that case the mass of
the new states is proportional to the medium mass density, and so
their production is no longer possible
\cite{Nelson:2007yq,Nelson:2008tn}. In our analysis therefore we will
not consider them.
\section{Vector and scalar signals}
\label{sec:vec-scalar}
In this section we discuss features of both, number of events and
recoil spectrum that can be used to differentiate vector from scalar
interactions signals. The discussion is split in light and effective
interactions. For the former we go into the ``deep'' light regime,
which from Figs.~\ref{fig:coh-limits} it can be seen corresponds to
$m_X\lesssim 10\;$MeV. For the latter, instead, we assume
$m_X\gtrsim 10^3\;$MeV. As for the bounds, in the light mediator case
we use limits from Figs. \ref{fig:coh-limits} while in the effective
case we rather use bounds in (\ref{eq:limits-effective}). In both
cases we use the $^8$B solar neutrino flux and xenon as target
material. This choice is well motivated by xenon-based experiments
such as XENONnT, LZ and DARWIN which will be subject to $^8$B neutrino
backgrounds \cite{Aprile:2015uzo,Akerib:2018lyp,Aalbers:2016jon}. In
order to be the less experiment dependent, we assume a ton-year
exposure and $100\%$ efficiency. For a given parameter choice results
are expressed in terms of
\begin{equation}
  \label{eq:R-param}
  R\equiv \frac{N_\text{SM+X}}{N_\text{SM}}\ , 
\end{equation}
where $N_\text{SM}$ refers to the number of events as expected in the
SM and depending on the case $X=V,S$.

\begin{figure*}
  \centering
  \includegraphics[scale=0.37]{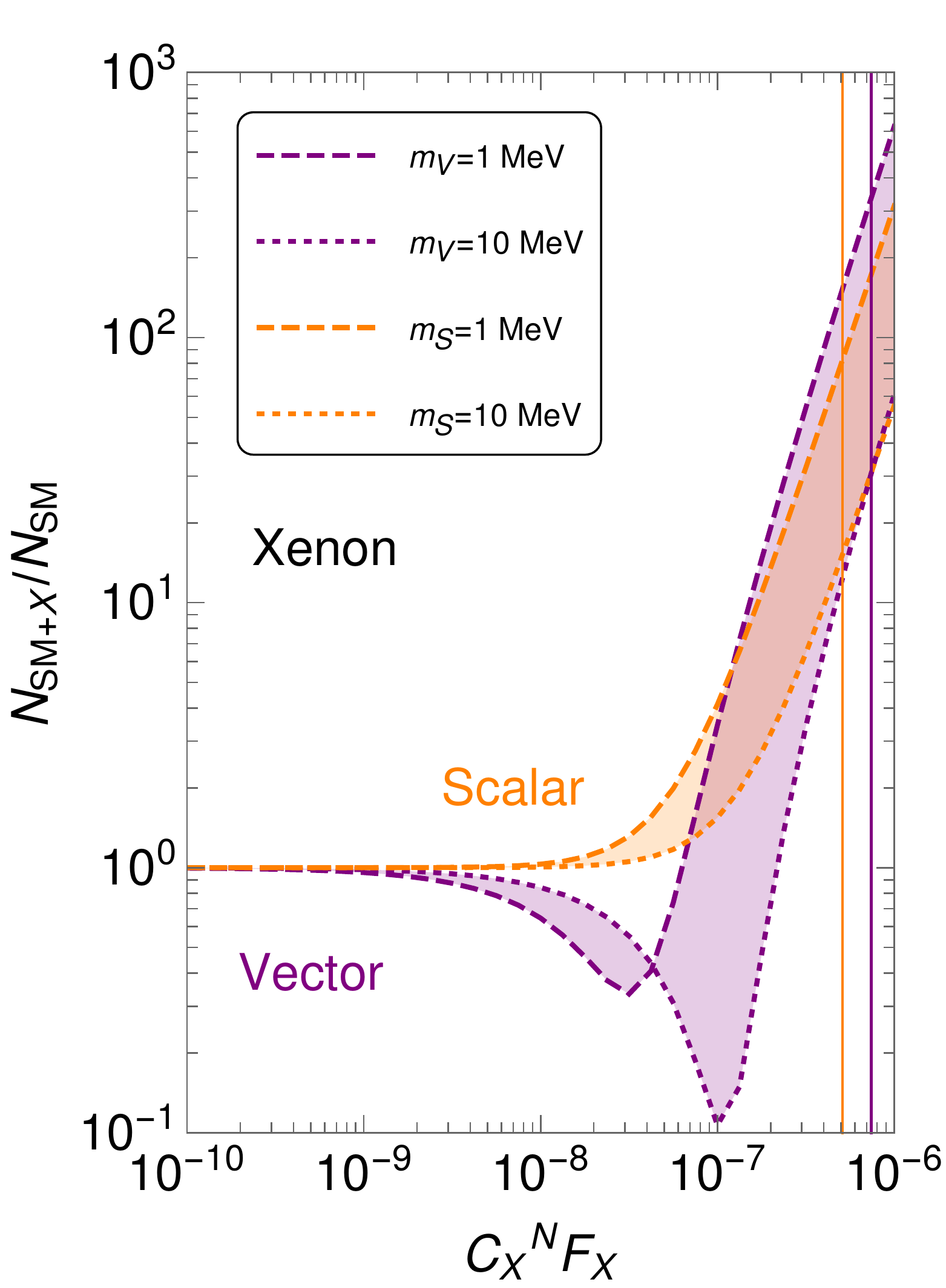}
  \hfill
  \includegraphics[scale=0.37]{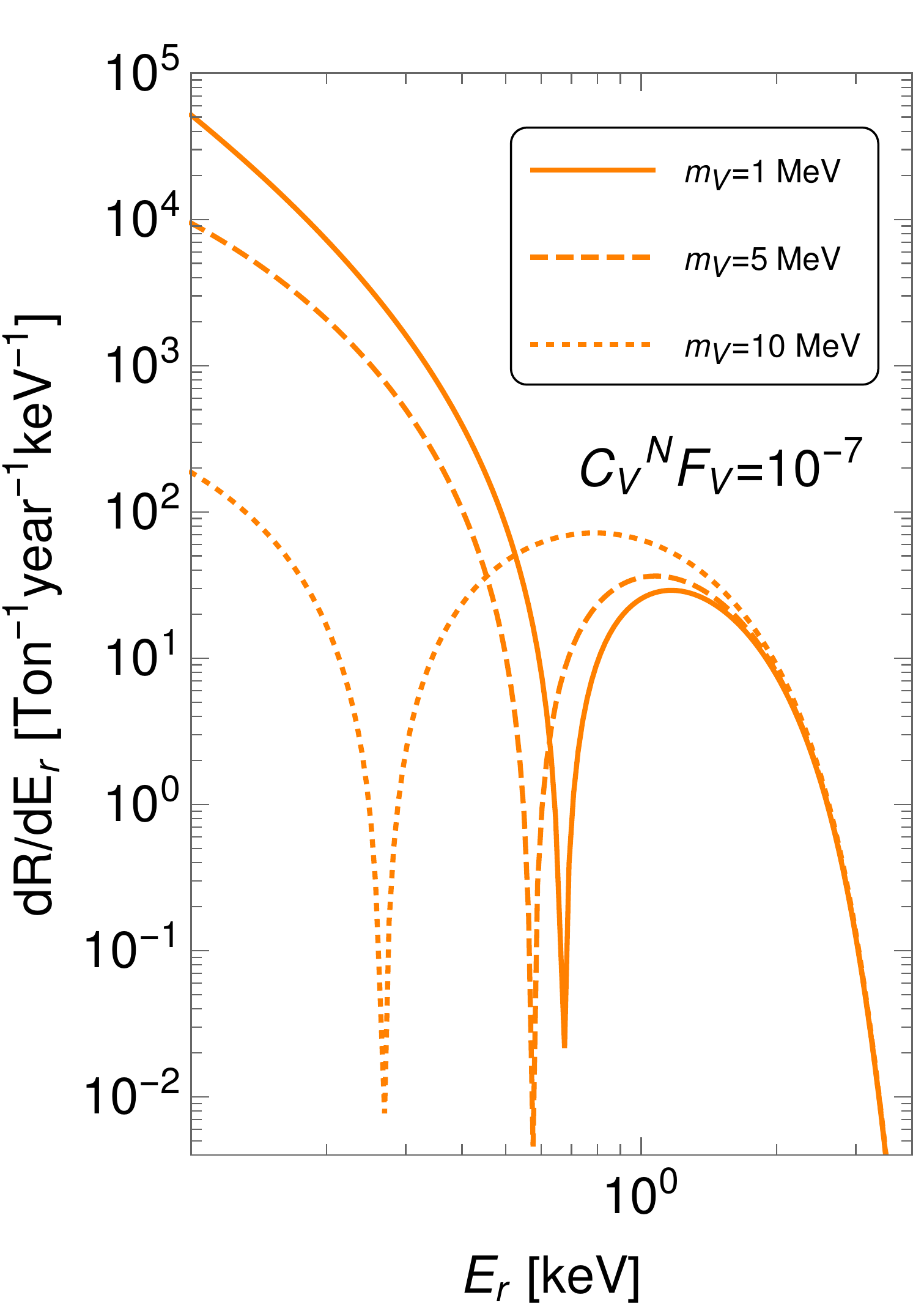}
  \hfill
  \includegraphics[scale=0.37]{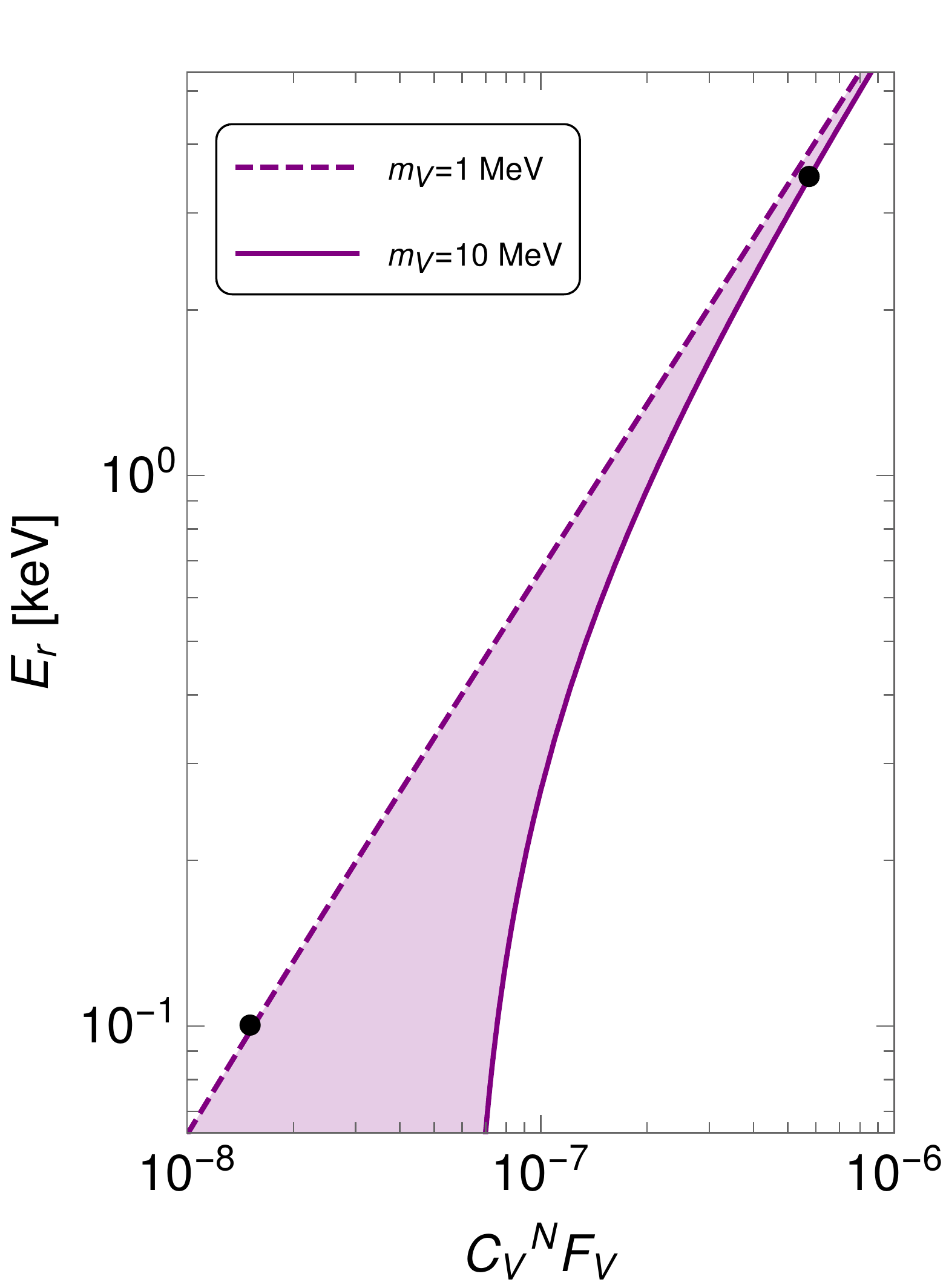}
  \caption{\textbf{Left graph}: Number of expected events normalized
    to the SM expectation for light vector (purple) and scalar
    (orange) mediators in a xenon detector assuming a one ton-year
    exposure and $100\%$ detector efficiency. The result was obtained
    assuming a $10^{-1}\;$keV energy recoil threshold, $X=V,S$. The
    vertical orange and purple lines indicate COHERENT upper limits
    for scalar and vector couplings, respectively.  \textbf{Middle
      graph}: Recoil spectrum for different light vector mediator
    masses as a function of recoil energy. This result demonstrates
    the presence of dips in the vector spectrum, in contrast to the
    scalar case. \textbf{Right graph}: Energy recoil region where dips
    are found in the light vector mediator case. The black dots
    located at
    $(C_V^NF_V,E_r)=(1.5\times 10^{-8},10^{-1}\;\text{keV})$ and
    $(C_V^NF_V,E_r)=(5.8\times 10^{-7},3.5)\;\text{keV}$ fix the
    boundaries where dips, though present, are no longer observable in
    the recoil spectrum. See text (Sec. \ref{sec:LM-VplusS}) for
    further details.}
  \label{fig:events}
\end{figure*}
\subsection{Light mediators}
\label{sec:LM-VplusS}
For the light vector mediator case $C_V^NF_V\leq 7.4\times 10^{-7}$,
while for scalar $C_S^NF_S\leq 5.1\times 10^{-7}$. Assuming universal
quark couplings these bounds can be translated into bounds in xenon
just by scaling by $A_\text{Xe}/A_\text{Cs}\simeq 0.99$. For the
calculation of the recoil spectrum we vary $E_r$ from $10^{-1}\;$keV
up to the recoil energy allowed by the neutrino energy tail of the
$^8$B neutrino spectrum, $E_\nu=16.56\,$MeV \cite{Bahcall:2004pz}. For
the calculation of the event spectrum $N_\text{events}$ we use a
$10^{-1}\;$keV threshold and integrate up to $E_r=97\;$keV. With that
choice we get $N_\text{events}^\text{SM}=760.25$.

The expected number of events (normalized to the SM expectation) as a
function of $C_XF_X$ is shown in the left graph in
Fig. \ref{fig:events}. In there one can see that for small coupling,
$C_X^NF_X\lesssim 10^{-9}$, the new interactions do not generate any
sizable deviation above or below the SM expectation. Right after that
value vector interactions start depleting the SM contribution, while
effects of the scalar (which can only produce enhancements of the
signal) start being visible only above $10^{-8}$. From that point on
up to about an order of magnitude ($7\times 10^{-8}$) vector and
scalar interactions behave rather differently, producing event rate
spectra which have no overlapping at all. Thus in that coupling range,
no matter the mediator mass as far as it is light, measurements can
differentiate between them both. From that value and up to the value
allowed by COHERENT constraints (vertical orange and purple lines for
scalar and vector, respectively), three regions can be distinguished:
Region I where $R=N_{\text{SM}+X}/N_\text{SM}\gtrsim 84$, region II
with $R\subsetsim (1,84]$ and region III where $R<1$. Regions I and
III are covered only by vector interactions, therefore a signal
featuring such values will favor a light vector mediator over the
scalar. Region II is problematic in the sense that in there using only
measurements of $R$ one cannot tell the nature of the new
contribution. There is however an interesting way through which
measurements of $R$ combined with measurements of the recoil spectrum
can provide a conclusive answer in almost all parameter space. Let us
discuss this in more detail.

\begin{figure*}
  \centering
  \includegraphics[scale=0.372]{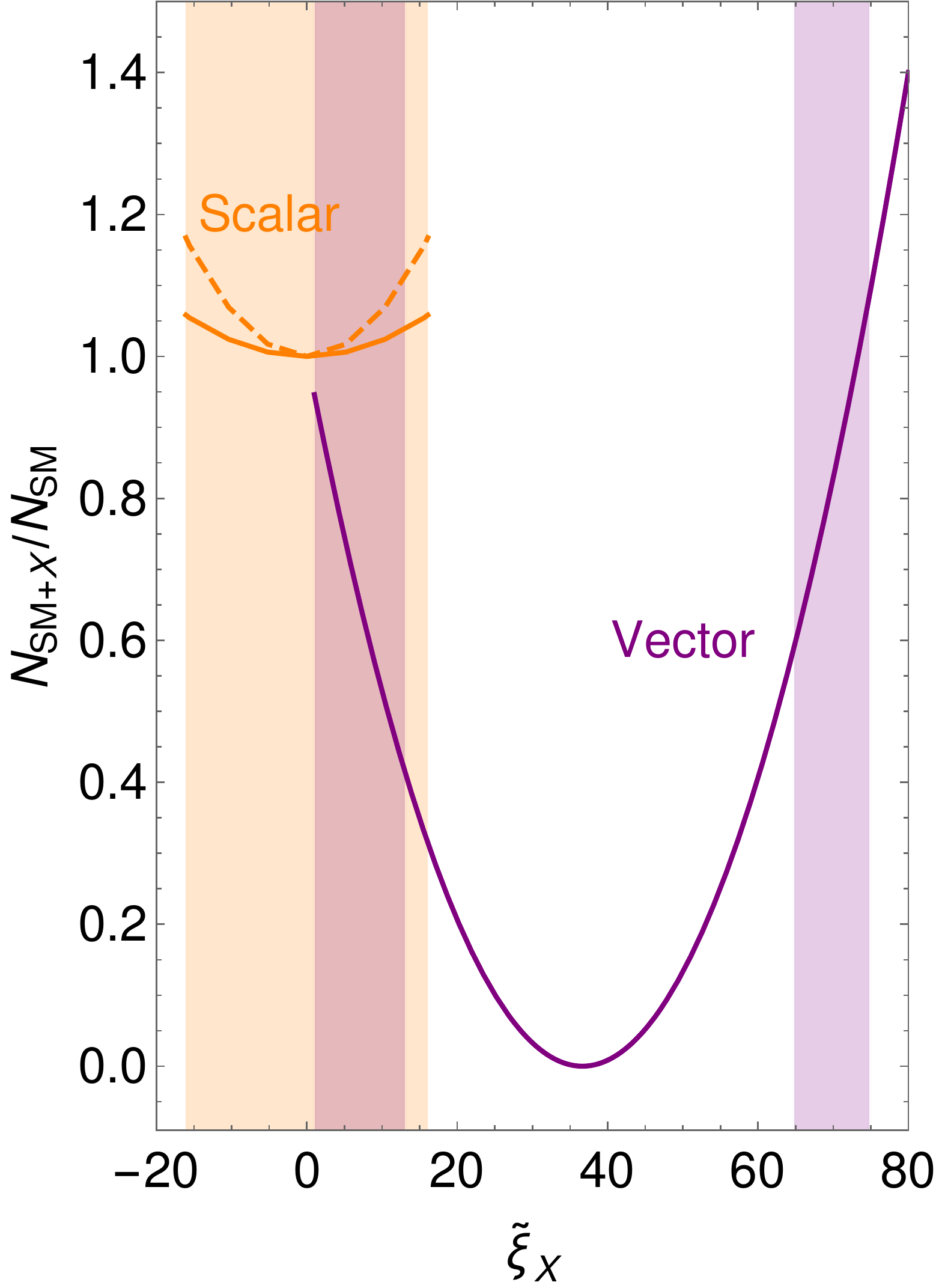}
  \hfill
  \includegraphics[scale=0.37]{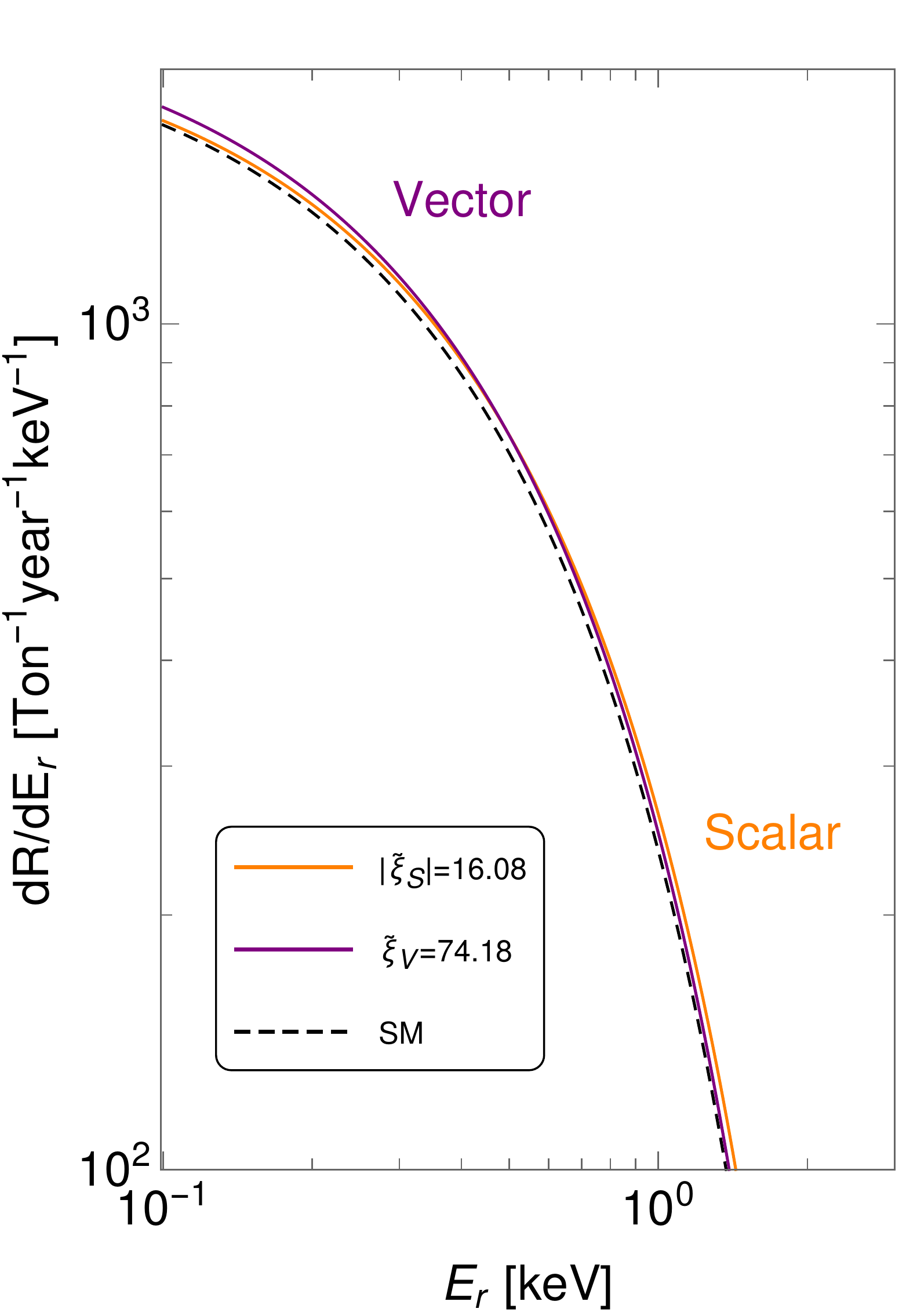}
  \hfill
  \includegraphics[scale=0.37]{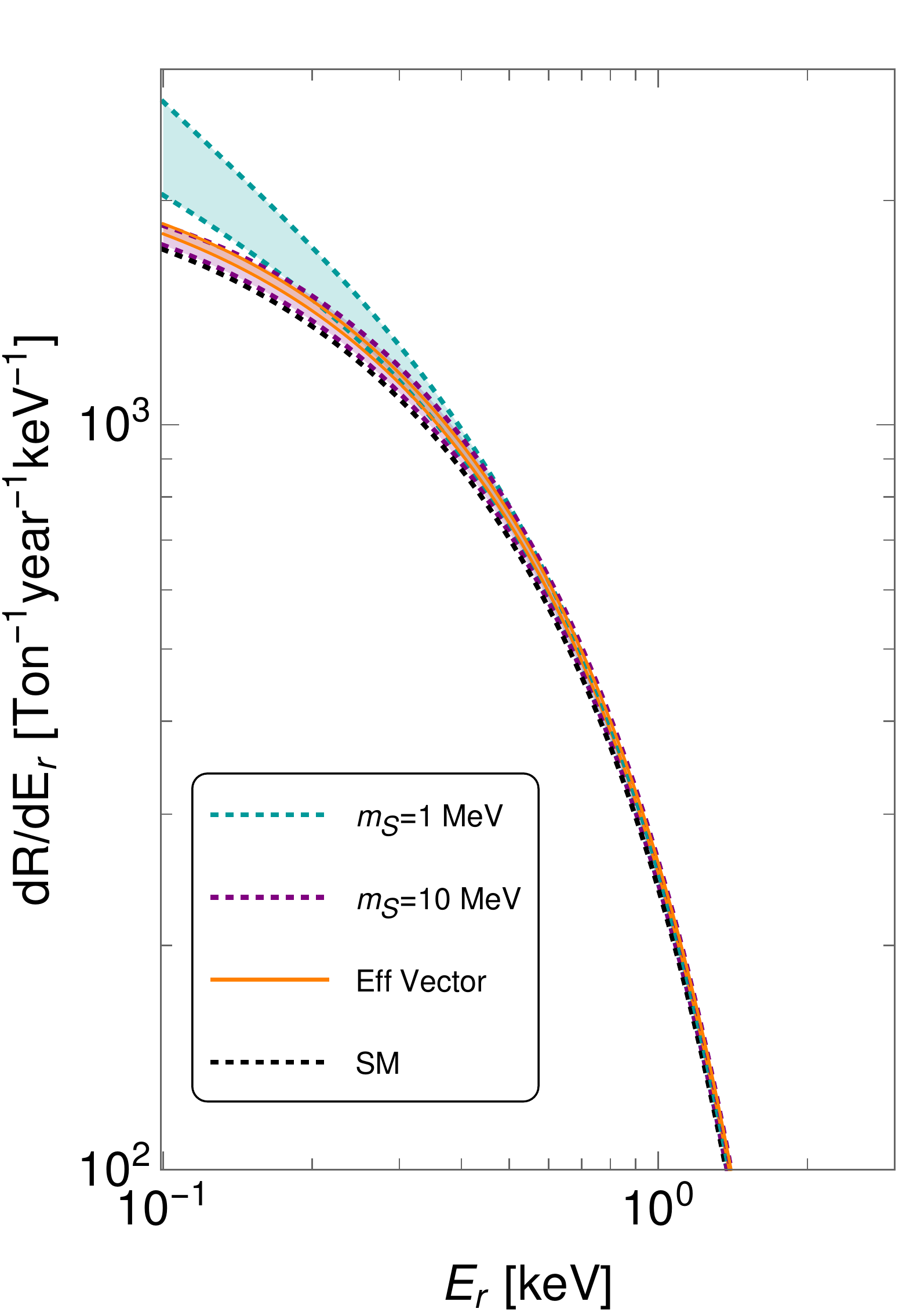}
  \caption{\textbf{Left graph}: Expected number of events in the
    presence of vector and scalar effective interactions
    ($m_X\gtrsim 10^3\;$MeV) normalized to the SM expectation as a
    function of the effective parameter
    $\widetilde{\xi}_X\;\; (X=V,S)$. The calculation is done assuming
    a $10^{-1}\;$keV threshold, a 1 ton-year exposure, 100\% detector
    efficiency and xenon as target material. Included as well is the
    result for scalar interactions obtained by taking a $1\,$keV
    threshold (dashed orange curve). The shadowed stripes indicate the
    $90\%$ CL limits from COHERENT data. \textbf{Middle graph}: Recoil
    spectra for which effective vector and scalar interactions cannot
    be differentiated by only measurements of
    $R\equiv N_\text{SM+X}/N_\text{SM}\subset (1.0,1.05]$.  This
    result demonstrates that measurements of $R$ combined with
    measurements of the recoil spectrum can be used to tell whether
    the signal is due to vector or scalar effective
    couplings. \textbf{Right graph}: Comparison of recoil spectra for
    light scalar mediators and effective vector
    interactions. Integration of these recoil spectra lead to
    $R\subset [1.05,1.08]$ in both cases. Combined measurements of $R$
    and the recoil spectrum can then be used to distinguish the
    interaction responsible for the signal. See text
    (Sec. \ref{sec:eff-int}) for further details.}
  \label{fig:effective-events}
\end{figure*}
In contrast to scalar interactions vector can produce depletions below
the SM expectation. And at the recoil spectrum level can lead to dips
in the spectrum, as exemplified in the middle graph in
Fig.~\ref{fig:events} which shows the recoil spectrum for
$m_V=1,5,10\;$MeV calculated for $C_VF_V=10^{-7}$. For a given
coupling $C_V^NF_V$ and vector boson mass $m_V$ the location of such
dip is determined by the condition \cite{AristizabalSierra:2019ufd}
\begin{equation}
  \label{eq:dips}
  E_r=\frac{C_V^NF_V-\sqrt{2}G_F|g_V|m_V^2}{2\sqrt{2}G_F|g_V|m_N}\ .
\end{equation}
For the parameters that define region II and within the ``deep'' light
vector mediator window we are considering this implies that those dips
are located within the recoil energy interval $[0.46,4.9]\;$keV, with
the left boundary of the interval obtained for
$(C_V^NF_V,m_V)=(7\times 10^{-8},1\;\text{MeV})$ and the right for
$(C_V^NF_V,m_V)=(7.4\times 10^{-7},4\;\text{MeV})$.  Bearing in mind
that the observable recoil energy window is defined by
$[10^{-1},3.5]\;$keV (the value to the left determined by rather
optimistic future thresholds, while the value to the right by the
kinematic endpoint energy of the $^8$B neutrino spectrum)
\footnote{Note that choosing a less optimistic threshold, which very
  likely will be the case, does not change our conclusion: observation
  of a dip will discard scalar interactions as being responsible for
  the signal.}, this means that some dips are not observable. This
happens for parameters in the range
$C_V^NF_V\lesssim 1.5\times 10^{-8}$ and
$C_V^NF_V\gtrsim 5.8\times 10^{-7}$, determined by the points where
the boundaries $10^{-1}\;$keV and $3.5\;$keV intercept the isocontours
$m_V=1\;$MeV and $m_V=10\;$MeV in the right graph in
Fig.~\ref{fig:events}, indicated by the black points. For
$C_V^NF_V\lesssim 1.5\times 10^{-8}$ the expected number of events
barely exceeds the SM expectation, and so in that regard that region
basically does not differ from the region for which
$C_V^NF_V\lesssim 10^{-9}$. For $C_V^NF_V\gtrsim 5.8\times 10^{-7}$,
once is already in a region where COHERENT bounds on scalar
interactions rules out the possibility of a signal from scalar
couplings; and so differentiation is possible. In conclusion, combined
measurements of $R$ and of the recoil spectrum will suffice---in
principle---to determine the nature of the new physics in region II as
well.
\subsection{Effective interactions}
\label{sec:eff-int}
For the effective vector mediator analysis we drop the $q^2$
dependence in (\ref{eq:xiV-xiS}) and write the coupling according to
$\xi_V=g_V+\widetilde{\xi}_V$, with $\widetilde{\xi}_V$ subject to the
constraints in (\ref{eq:limits-effective}). For the effective scalar
calculation we drop as well the $q^2$ dependence in $\xi_S$ in
(\ref{eq:xiV-xiS}) and treat $\xi_S$ as a free parameter subject to
the limits in (\ref{eq:limits-effective}). By comparing the SM+vector
and SM differential cross sections, Eqs. (\ref{eq:x-sec-V}), one can
see that in the effective vector case deviations in the CE$\nu$NS
process are entirely controlled by the ratio $\xi_V^2/g_V^2$
\cite{AristizabalSierra:2017joc}. In contrast scalar interactions have
a different energy dependence, and so a full calculation of the recoil
spectrum and its integration according to Eq. (\ref{eq:events}) are
required.

The left graph in Fig.~\ref{fig:effective-events} shows the expected
number of events (normalized to the SM expectation) as a function of
the effective couplings $\widetilde{\xi}_X\;\;(X=V,S)$. The shadowed
vertical stripes indicate the 90\% CL limits on $\xi_V$ (purple) and
$\xi_S$ (orange). In terms of $R$ two regions can be identified. A
region entirely dominated by the vector interaction where
$R\lesssim 0.94$ (the new vector interaction destructively interferes
with the SM contribution) and a second region where $R>1$. In the
latter two subregions can be identified: (i) One where $R$ exceeds
$1.05$ but never above $1.08$, (ii) a second where $R$ goes above the
SM expectation but does not exceed $1.05$. Measurements yielding
$R\lesssim 0.94$ or $R\subset [1.05,1.08]$ will point to a vector
interaction as responsible for the signal. In contrast, for
measurements resulting in $R\subset (1,1.05]$, $R$ alone cannot be
used to disentangle whether the signal is related with a vector or a
scalar interaction.

To break such ``degeneracy'' one can check the recoil energy spectrum
and see whether vector and scalar interactions lead to different
spectral features. The key observation here is that the differential
cross sections for scalar and vector interactions have different
recoil energy dependencies. As a consequence the effect of vector
interactions is just an overall rescaling of the SM expectation, while
for the scalar coupling energy dependent differences are found. This
can be seen in the middle graph in Fig. \ref{fig:effective-events}
which shows the vector and scalar recoil spectra evaluated for
couplings in that region: $\xi_V=74.18$ and $|\xi_S|=16.08$. One can
see that scalar interactions produce recoil spectra that at small
recoil energies tend to overlap with the SM expectation and to
departure from it at higher $E_r$. This is not the case for
vector-induced spectra which follow the SM spectrum up to a
multiplicative factor. Disentanglement of vector and scalar
interactions in the region $R\subsetsim (1.0,1.05]$ can then be
done---in principle---combining information on $R$ and on the recoil
spectrum, provided the new physics couplings have values such that
sizable departures from the SM prediction are observed and the
detector has a good spectral resolution.

Regarding the effective interactions analysis there is however a
caveat on some of our conclusions. Measurements yielding
$R\subsetsim [1.05,1.08]$ can be obtained as well in the light scalar
mediator case, as shown in the left graph in Fig. \ref{fig:events}. So
although such measurements cannot result from effective scalar
interactions, they can if the scalar mediator is light. For a light
scalar, $R\subsetsim [1.05,1.08]$ results from
$C_S^NF_S\subset [1.2,2.6]\times 10^{-8}$ for $m_V=1\,$MeV or
$C_S^NF_S\subset [1.5,3.7]\times 10^{-8}$ for $m_V=10\,$MeV. This
ambiguity can be---in principle---removed with the aid of the recoil
spectrum. As shown in the right graph in
Fig.~\ref{fig:effective-events}, whether this is the case depends on
$m_S$. For values of $m_S$ close to 1 MeV the recoil spectrum is
rather peaked at low energies and largely differs from the recoil
spectra induced by effective vector interactions. As $m_S$ increases
towards values close to 10 MeV a strong overlapping between light and
effective spectra is instead found. Thus the question of whether one
can disentangle vector (effective) and scalar (light) interactions in
the region $R\subsetsim [1.05,1.08]$ depends to a large extent on
$m_S$.

\begin{figure*}
  \centering
  \includegraphics[scale=0.37]{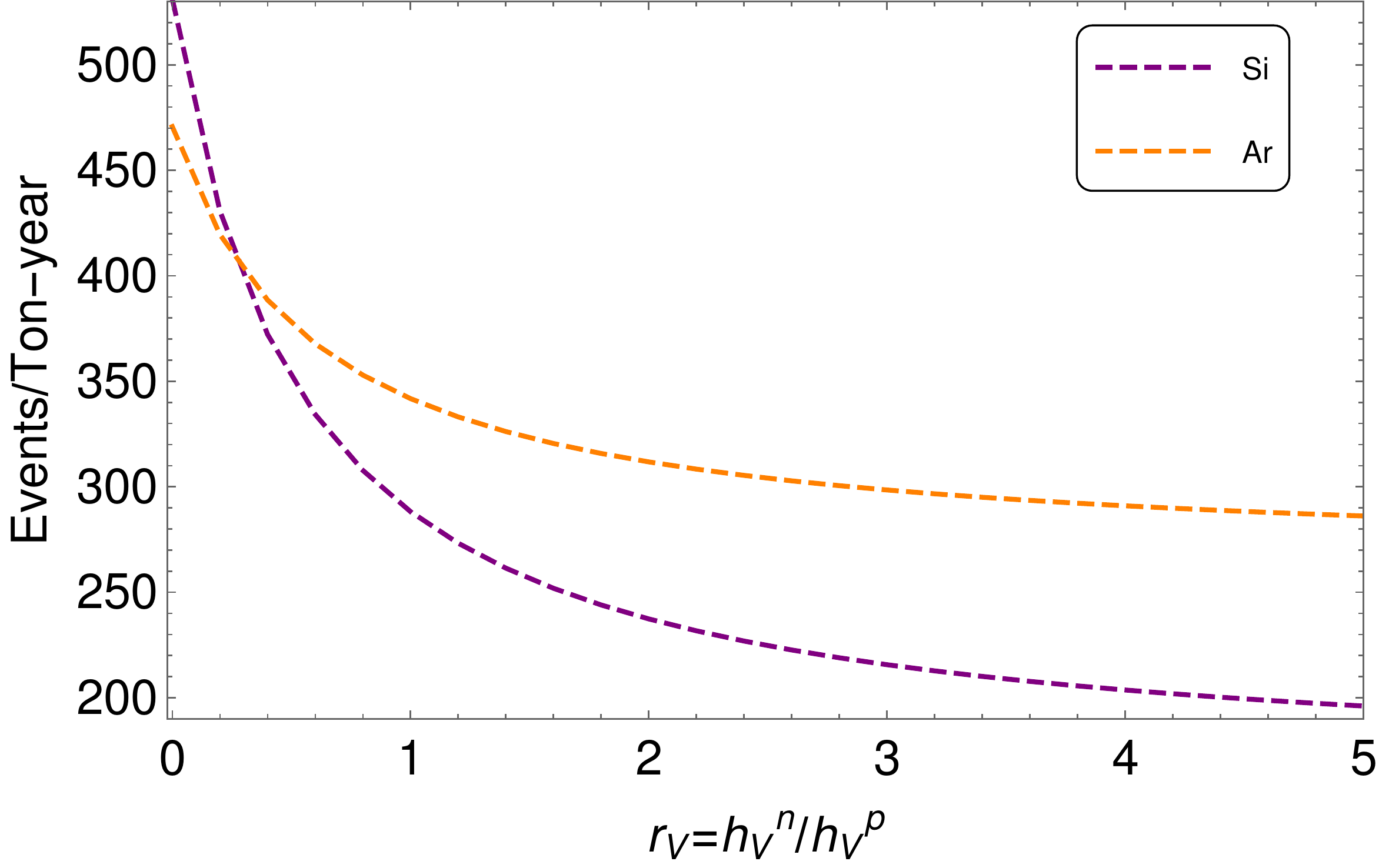}
  \hfill
  \includegraphics[scale=0.37]{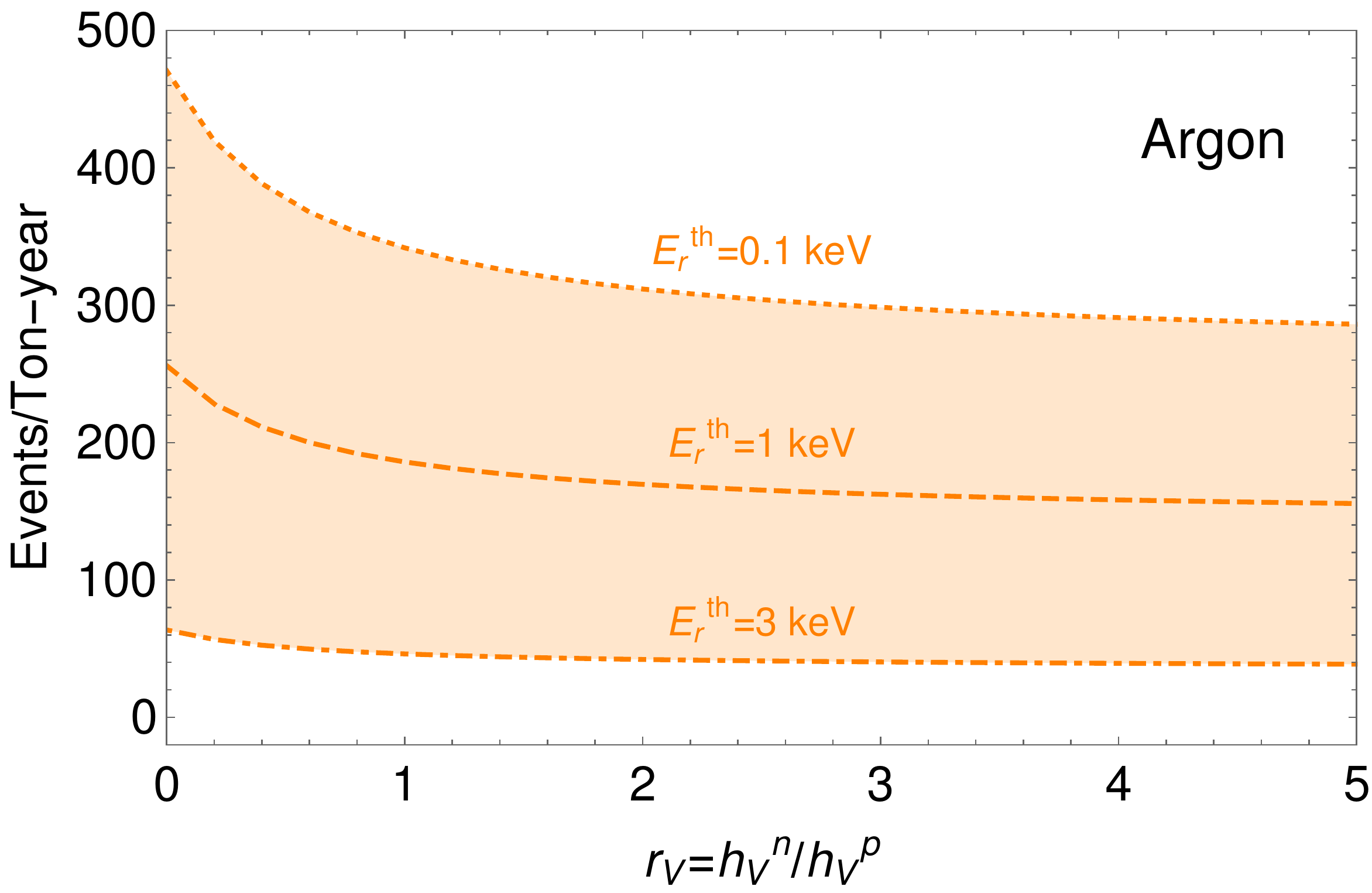}
  \caption{\textbf{Left graph}: Expected number of events in silicon
    and argon detectors as a function of $r_V=h_V^n/h_V^p$ ($h_V^n$
    and $h_V^p$ refer to the couplings of the new ``heavy'' vector to
    neutrons and protons respectively). The result was obtained
    assuming a xenon detector has measured $R=1.08$ (i.e.
    $N_\text{events}^\text{Xe}=821\; \text{events/ton-year}$) and an
    effective vector interaction has been established as responsible
    for the signal (see sec. \ref{sec:eff-int}). For this particular
    benchmark case, measurements $N_\text{events}^\text{Si}\simeq 288$
    and $N_\text{events}^\text{Ar}\simeq 342$ will favor an isospin
    conserving interaction ($r_V=1$), deviations from these values
    will instead favor isospin violation. \textbf{Right graph}: Number
    of events/ton-year in an argon detector for various recoil energy
    thresholds. For $E_r^\text{th}>3\,$keV, the lack of statistics
    will make the isospin test analysis hard. In that case exposures
    above 1 ton-year will be required \cite{Aalseth:2017fik}.}
  \label{fig:ici-vs-ivi}
\end{figure*}
\section{Isospin conserving versus isospin violating interactions}
\label{sec:isospin-cons-viol}
In what follows we discuss the capability of ton-size DM detectors to
identify the isospin nature of the new interaction. From Eqs.
(\ref{eq:fundamental-couplings-nuclear-couplings-vector}) and
(\ref{eq:fundamental-couplings-nuclear-couplings-scalar}) one can see
that the conditions 
\begin{align}
  \label{eq:isospin-conserving}
  h_V^u=h_V^d\ ,\qquad h_S^u=11.9\,h_S^d\ ,
\end{align}
assure isospin conserving interactions (for the scalar case we are
using the hadronic form factors central values, see
Eq. (\ref{eq:ftqn})). Deviations from these relations lead to isospin
violation, of which one can distinguish three (extreme) particular
cases. Protophobic and neutrophobic interactions and degeneracies. The
latter defined as a region in parameter space for which the new
physics signal exactly vanishes in a given detector (for a given
nuclide). The relations between the fundamental quark couplings that
define each case depend---of course---on whether the interaction is
vector or scalar (regardless of the size of the mediator
mass). Starting with the vector case one finds (degeneracy in a
$(A_1,Z_1)$ nucleus):
\begin{align}
  \label{isospin-cases-vec}
  &\text{Protophobic}:\quad h^u_V=-h^d_V/2\ ,
    \nonumber\\
  &\text{Neutrophobic}:\quad h^u_V=-2h^d_V\ ,
    \nonumber\\
  &\text{Degeneracy}:\quad
    h_V^u=-\frac{2A_1-Z_1}{A_1+Z_1} h_V^d\ .
\end{align}
For scalar interactions instead the relations can be written as
\begin{alignat}{2}
  \label{isospin-cases-sca}
  &\text{Protophobic}:\quad& h^u_S=&-\frac{m_u}{m_d}
    \frac{f_{T_d}^p}{f_{T_u}^p}h^d_S=-0.92 h^d_S\ ,
    \nonumber\\
  &\text{Neutrophobic}:\quad& h^u_S=&-\frac{m_u}{m_d}
    \frac{f_{T_d}^n}{f_{T_u}^n}h^d_S=-1.11 h^d_S\ ,
    \nonumber\\
  &\text{Degeneracy}:\quad&
    h_S^u=&-\frac{m_u}{m_d}
    \frac{Z_1m_p f_{T_d}^p + N_1 m_n f_{T_d}^n}
    {Z_1m_p f_{T_u}^p + N_1 m_n f_{T_u}^n} h_S^d
    \nonumber\\
    &\quad&=&-\frac{9.01 A_1 - 0.81 Z_1}{8.07 A_1 + 0.79 Z_1}h_S^d\ .
\end{alignat}
Experimentally establishing isospin conservation or these
isospin-violating scenarios (or any other intermediate
isospin-violating case) cannot be done on the basis of a single
measurement. In the case of degeneracies is rather obvious that at
least two independent measurements are required. One in which no
deviation from the SM is observed and one where degeneracy is broken,
at least partially. In the protophobic and neutrophobic cases the
reason as well is rather simple. The new physics is controlled by
$C_X^NF_X$ which can be ``reparametrized'' according to
\begin{align}
  \label{eq:reparametrization-1}
  F_VC_V^N=F_Vh_V^p\left[Z + r_V(A-Z)\right]\ ,
  \\
  \label{eq:reparametrization-2}
  F_SC_S^N=F_Sh_S^p\left[Z + r_S(A-Z)\right]\ .
\end{align}
These quantities involve the overall factor and the ratio
$r_X=h_X^n/h_X^p$, which a single measurement cannot fix without
ambiguity. Two measurements in different detectors instead will enable
pinning down their values, at least in the effective case.

To demonstrate how this can be done we focus on the effective vector
case and assume that out of the two different measurements one is done
in a xenon-based detector. This means that a deviation in the SM
prediction has been observed and that using the arguments of the
previous section the effective vector nature of the new interaction
has been established. Therefore measurement of $R$ in the xenon
detector fixes $\tilde\xi_V=\tilde\xi|_\text{Xe}$. This information
can be used to fix the overall factor $F_Vh^p_V$ to
$F_Vh^p_V=\tilde\xi|_\text{Xe}/(Z_\text{Xe} + N_\text{Xe}r_V)$. With
that information as an input one can then calculate the number of
events in a second detector (D2) as a function of $r_V$. The value of
$r_V$ for which $N_\text{theory}^\text{D2}=N_\text{Exp}^\text{D2}$
will tell whether the new interaction is isospin conserving or
violating: $r_V=1$ will establish isospin conservation, while any
other case will prove otherwise.  Note that the protophobic and
neutrophic scenarios will be favored by $r_V\ll 1$ and $r_V\gg 1$,
respectively.

\begin{figure*}
  \centering
  \includegraphics[scale=0.37]{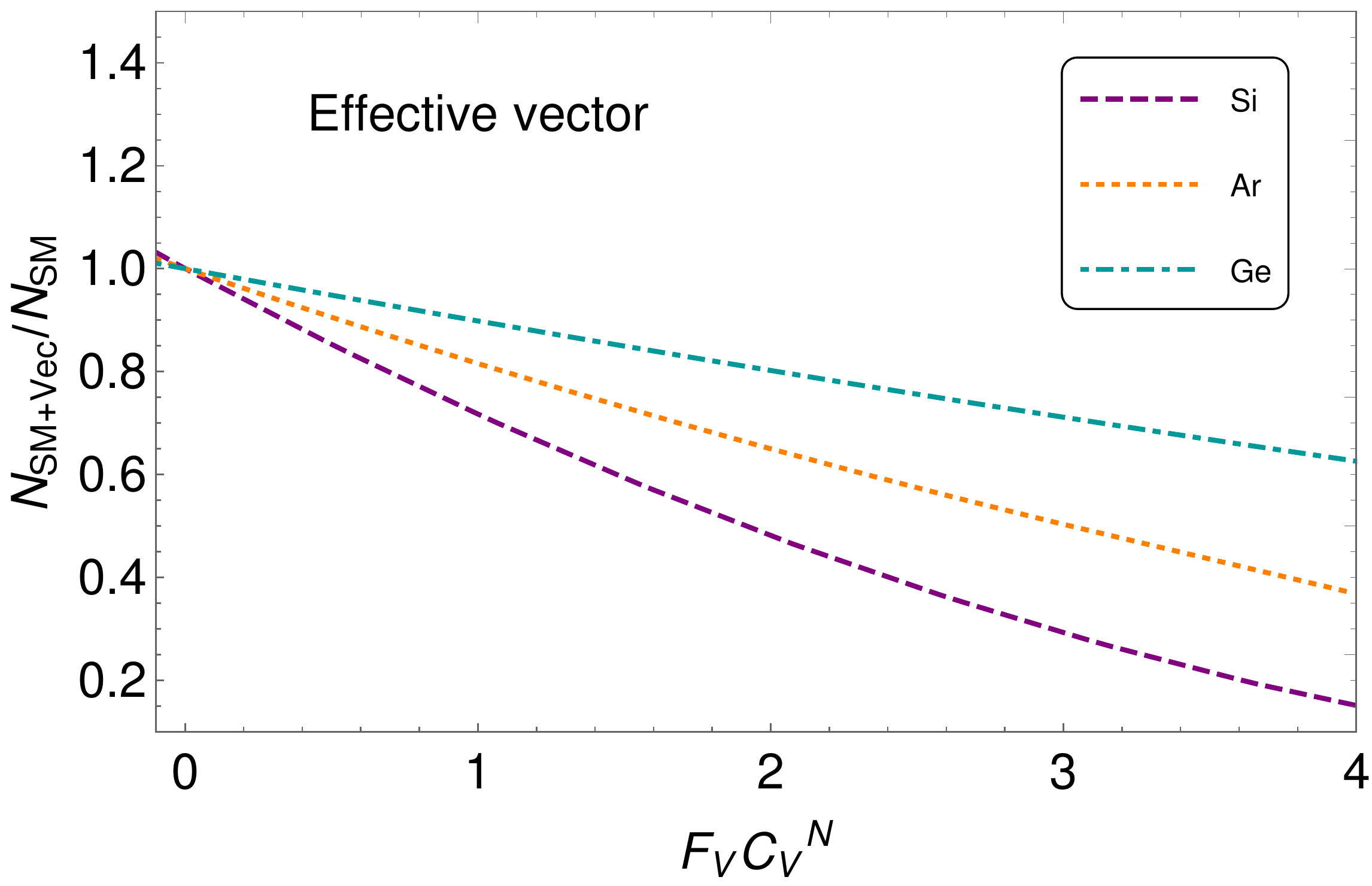}
  \hfill
  \includegraphics[scale=0.37]{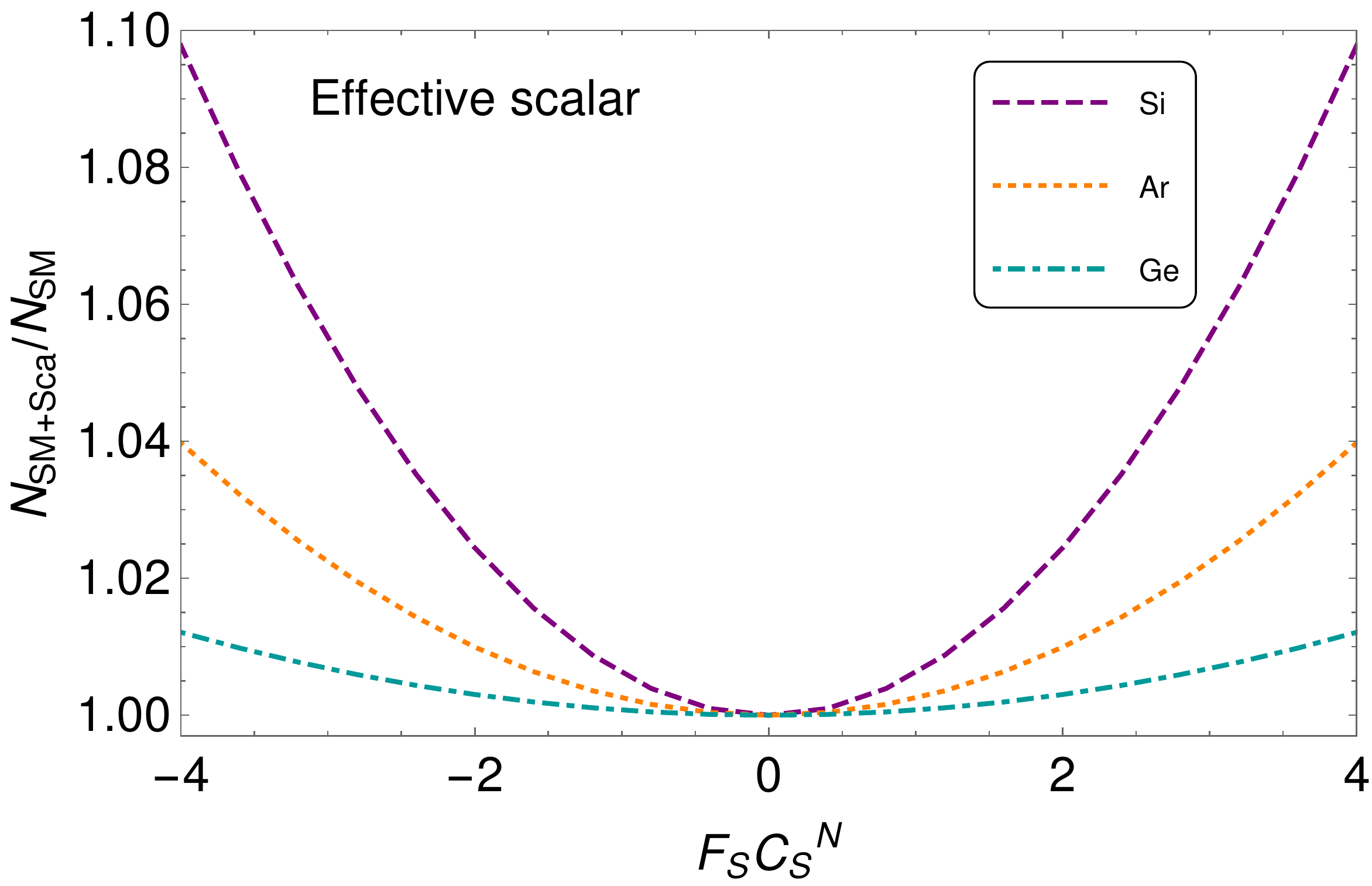}\\
  \includegraphics[scale=0.37]{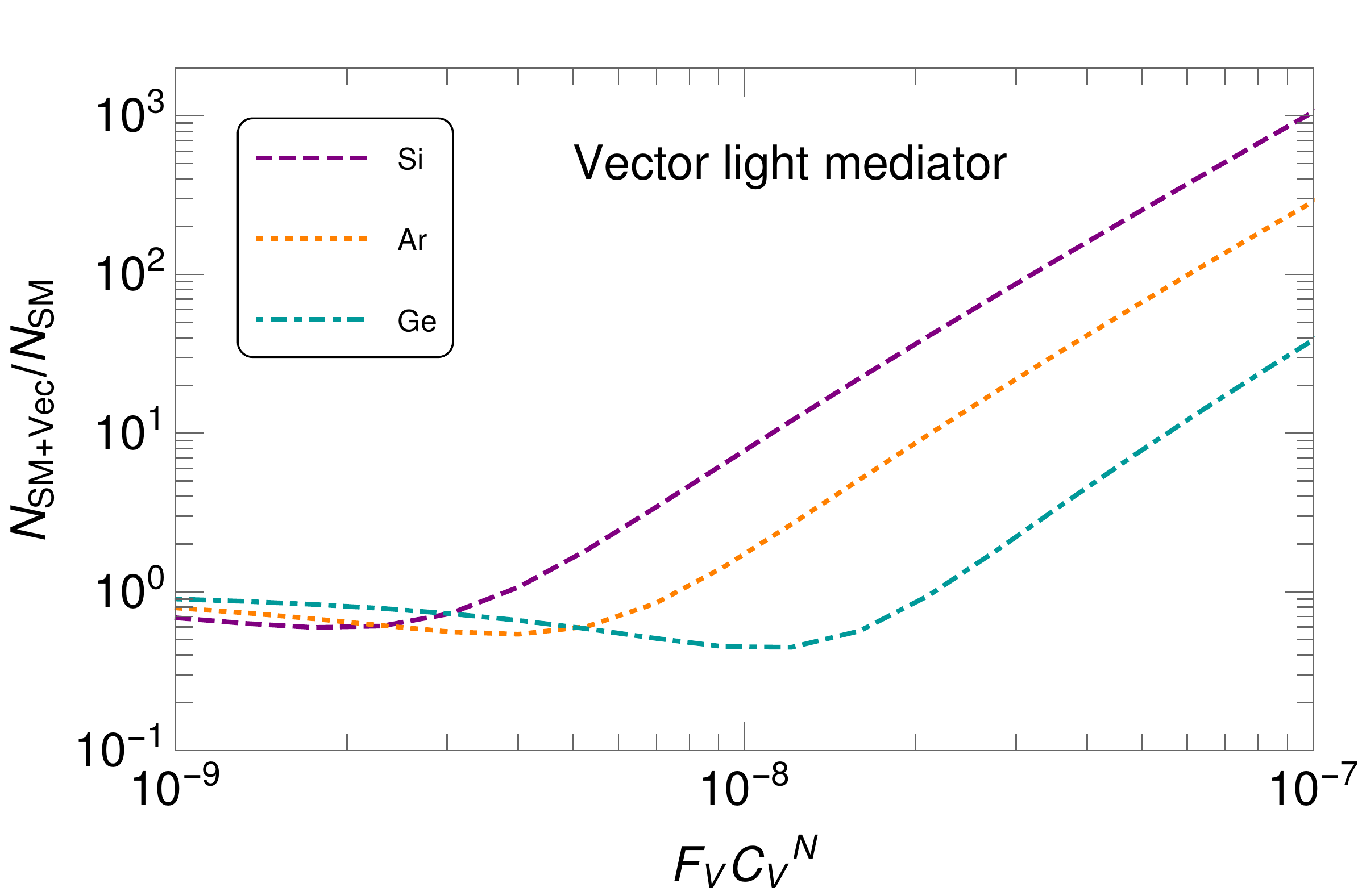}
  \hfill
  \includegraphics[scale=0.37]{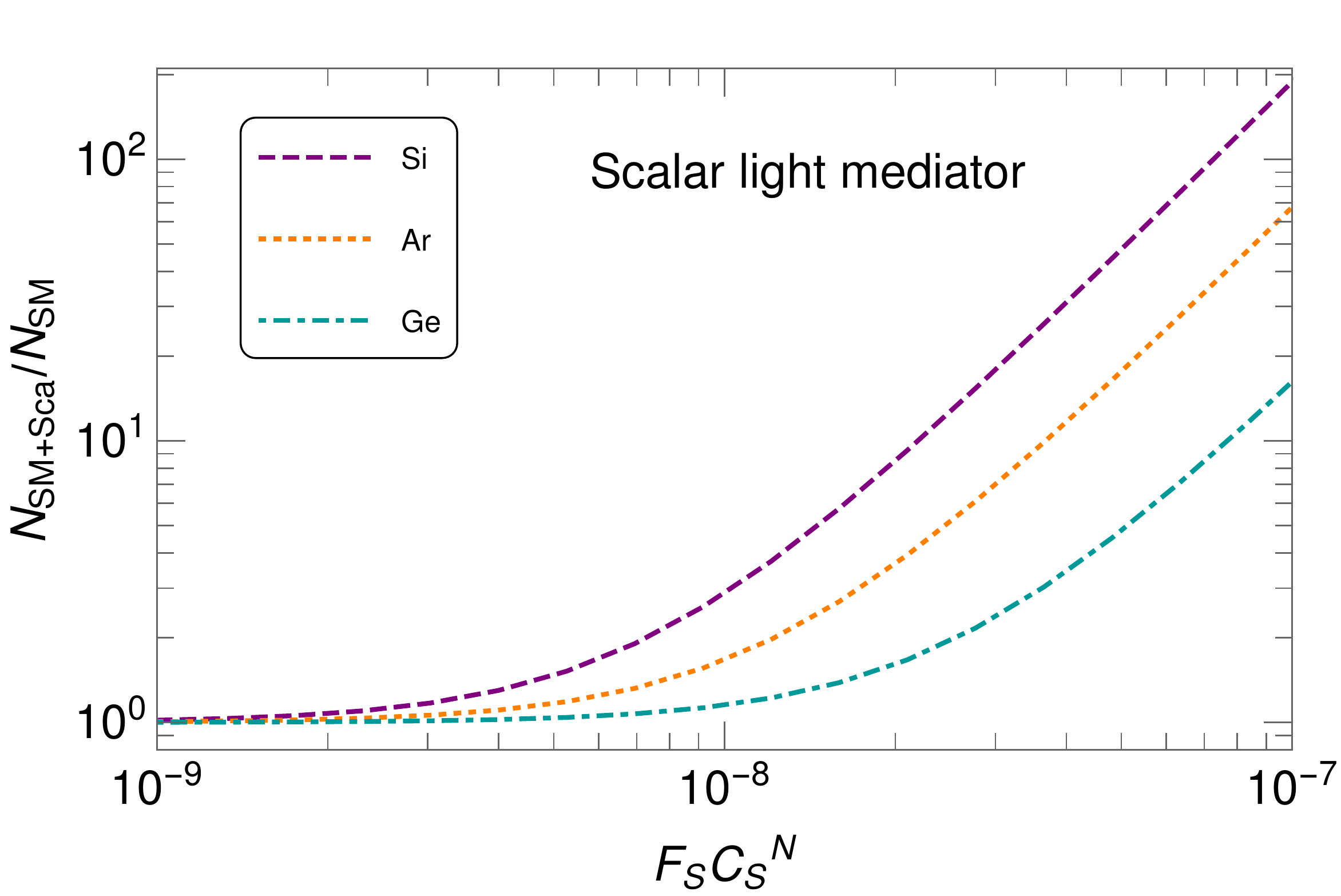}
  \caption{\textbf{Top-left graph}: Number of events normalized to the
    SM expectation in silicon, argon and germanium detectors assuming
    parameter space degeneracy in a xenon detector
    $C_X^N|_\text{Xe}=0$. The result was obtained assuming
    $E_r^\text{th}=0.1\;$keV and an effective vector
    interaction. \textbf{Top-right graph} Same as in the top-left
    graph but for effective scalar interactions. Both results
    demonstrate that the silicon detector performs better for
    degeneracy removal and allow to distinguish in almost all
    parameter space vector from scalar signals. \textbf{Bottom-left
      graph}: Same as in the top-left graph but for vector light
    mediators. As in the effective case---in general---silicon
    performs better. However in contrast to the effective case the
    signal is more degraded (below the SM) in germanium, so whether
    silicon or germanium are more suited for breaking xenon
    degeneracies depends on where the values of the parameters fall
    in. \textbf{Bottom-right graph}: Same as in the top-left graph but
    for scalar light mediators. For light mediators scenarios the
    calculation has been done fixing $m_X=1\;$MeV, value for which
    effects are maximized (within the mass range we are considering).}
  \label{fig:degeneracy}
\end{figure*}
To show the performance of different detectors we take the parameter
space point for which $R=1.08$, obtained for
$\tilde\xi_V|_\text{Xe}=74.77$ (see left graph in
Fig. \ref{fig:effective-events}). With the overall factor in
(\ref{eq:reparametrization-1}) fixed according to this number, we then
calculate the expected number of events in argon and silicon detectors
as a function of $r_V$. The result is displayed in the left graph in
Fig. \ref{fig:ici-vs-ivi}, which shows the number of events in each
detector as a function of $r_V$. Although done for a particular point
in parameter space, this result allows to capture the general
picture. Light isotopes have a stronger $r_V$ dependence, so are
better suited to determine the isospin character of the new
interaction. For silicon and for $r_V$ varying as shown in
Fig. \ref{fig:ici-vs-ivi} (left graph), $N_\text{events}$ decreases
about $63\%$ while for argon about $39\%$ (we checked for germanium as
well and in that case the variation is smaller). Note that this
conclusion is inline with analysis done in the context of isospin
violating DM \cite{Kelso:2017gib}. The number of events increases in
the neutrophobic case and decreases in protophobic scenarios. Such
behavior can therefore be used to identify the scenario to which the
new physics belongs. The results displayed in the left graph in
Fig.~\ref{fig:ici-vs-ivi} are derived by assuming a 0.1 keV
threshold. If one takes instead more realistic (experimentally)
values, the number of events can drastically decrease as shown in the
right graph in Fig.~\ref{fig:ici-vs-ivi}. In such a case larger
exposures will be required so to determine the isospin nature of the
new interaction, something feasibale at e.g. the Argo detector of the
Global Argon Dark Matter Collaboration \cite{Aalseth:2017fik}.

A second measurement then fixes $r_V$---of course within experimental
uncertainties---and that information can be used to partially
reconstruct the parameter space $h_V^{u,d}$ of the model responsible
for the signal. If it turns out that the model is protophobic or
neutrophobic partial reconstruction can be done right away with the
aid of Eqs. (\ref{isospin-cases-vec}). It is worth pointing out that
the procedure outlined here applies in the same manner to the
effective scalar case. It applies as well in light mediator scenarios
but in those cases $r_X$ is determined within a range, as we now
explain. Measurement in the xenon detector provides $R=R|_\text{Xe}$
and fixes the nature of the new interaction to, say, light scalar
mediator. This value for $R$ is obtained not for a single value of
$F_SC_S^N$, but within the interval
$\mathcal{I}=[F_SC_S^N|_\text{min},F_SC_S^N|_\text{max}]$ (see left
graph in Fig. \ref{fig:events}). For the overall factor in
Eq. (\ref{eq:reparametrization-2}) this translates into
$F_Sh_S^p=\mathcal{I}/(Z_\text{Xe}+N_\text{Xe}r_S)$, thus leading to
an spread in the calculation of $N_\text{theory}^\text{D2}$ in terms
of $r_s$.
\subsection{The case of degeneracies}
\label{sec:degeneracies}
We now turn to the discussion of parameter space degeneracies. For
that aim we will assume that degeneracy happens in a xenon detector
and will determine which among the silicon, argon and germanium
detectors performs better at breaking the degeneracy. For such choice
and according to Eqs.~(\ref{isospin-cases-vec}) and
(\ref{isospin-cases-sca}) the proton and scalar couplings can be
entirely expressed in terms of $h_X^d$
\begin{alignat}{2}
  \label{eq:proton-neutron-couplings-deg}
  &\text{Vector}:\quad h_V^p&=-1.252\;h_V^d\ , \quad h_V^n&=0.874\;h_V^d\ ,
  \nonumber\\
  &\text{Scalar}:\quad h_S^p&=-0.964\;h_S^d\ , \quad h_S^n&=0.672\;h_S^d\ .
\end{alignat}
With these results one can evaluate the nuclear coupling in silicon,
argon and germanium in terms only of $h_X^d$
\begin{alignat}{3}
  \label{eq:proton-neutron-couplings-deg}
  C_V^N|_\text{Si}&=-5.2 h_V^d\ , 
  \; C_V^N|_\text{Ar}&=-3.3 h_V^d\ ,
  \; C_V^N|_\text{Ge}&=-4.5 h_V^d\ ,
  \nonumber\\
  C_S^N|_\text{Si}&=-4.0 h_S^d\ ,
  \; C_S^N|_\text{Ar}&=-2.5 h_S^d\ ,
  \; C_S^N|_\text{Ge}&=-3.5 h_S^d\ .
\end{alignat}
For the calculation of the number of events in the different detectors
and for the effective case we then vary $F_Vh_V^d$ within the interval
$[-1,0]$ to assure $\xi_V>0$, as required by
(\ref{eq:limits-effective}). $F_Sh_S^d$ we instead vary within
$[-1,1]$. For light mediators we vary $h_X^{d}$ within
$[10^{-9},10^{-7}]$, to---again---guarantee constraints from COHERENT
are satisfied. We calculate as well the SM expectation for each
isotope:
\begin{equation}
  \label{eq:SM-expectation}
  N_\text{Event}^\text{SM}|_\text{Si}=136.5\ ,
  \;
  N_\text{Event}^\text{SM}|_\text{Ar}=234.5\ ,
  \;
  N_\text{Event}^\text{SM}|_\text{Ge}=417.8\ .
\end{equation}
The results are shown in Fig.~\ref{fig:degeneracy}. From the graphs on
top, which correspond to the effective vector (left) and scalar
(right) cases, one can see that silicon performs better than argon and
germanium. The vector interaction depletes $N_\text{Event}$ below the
SM expectation. For silicon the depletion amounts to about
$0.1\times N_\text{Event}^\text{SM}|_\text{Si}$.  For scalar
interactions deviations from the SM prediction are somehow more
modest, in silicon they amount at most to $10\%$. Thus, breaking
scalar degeneracies in xenon will require a silicon detector with a
high event rate resolution. All in all, among the isotopes we are
considering, study of degeneracies in xenon generated by effective
interactions should be---ideally---done combining measurements in
xenon and silicon detectors.

From the graphs on top one can see as well that identification of
vector and scalar interactions can be done in all the parameter space
we are considering just by measurements of $R$. Depletions below the
SM expectation will establish vector interactions as responsible for
the signal. Enhancements, instead, will favor scalar interactions.
Measurements in a silicon detector will therefore not only remove the
parameter space degeneracy but will also establish the nature of the
new physics.

\begin{figure*}
  \centering
  \includegraphics[scale=0.36]{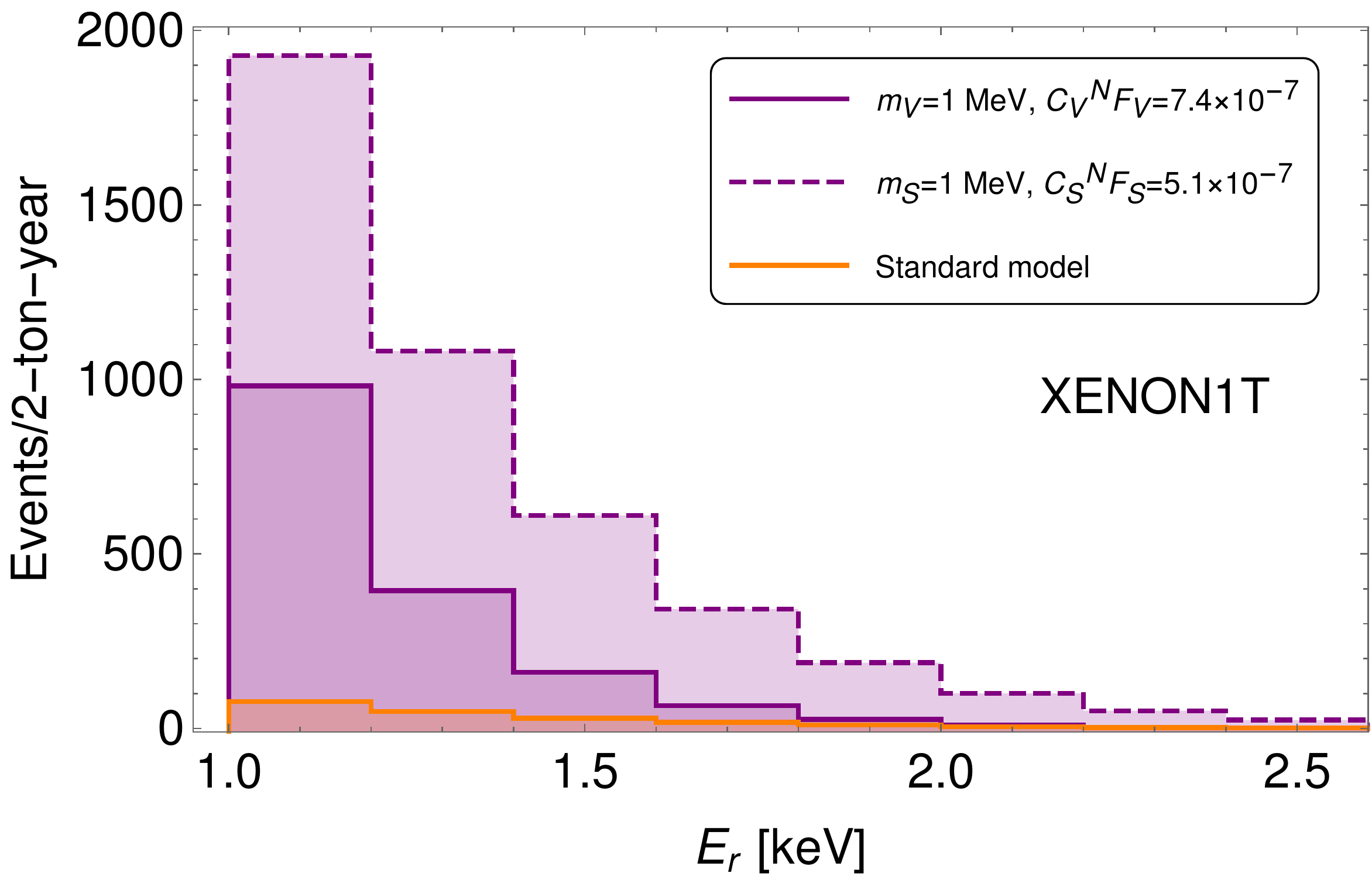}
  \hfill
  \includegraphics[scale=0.36]{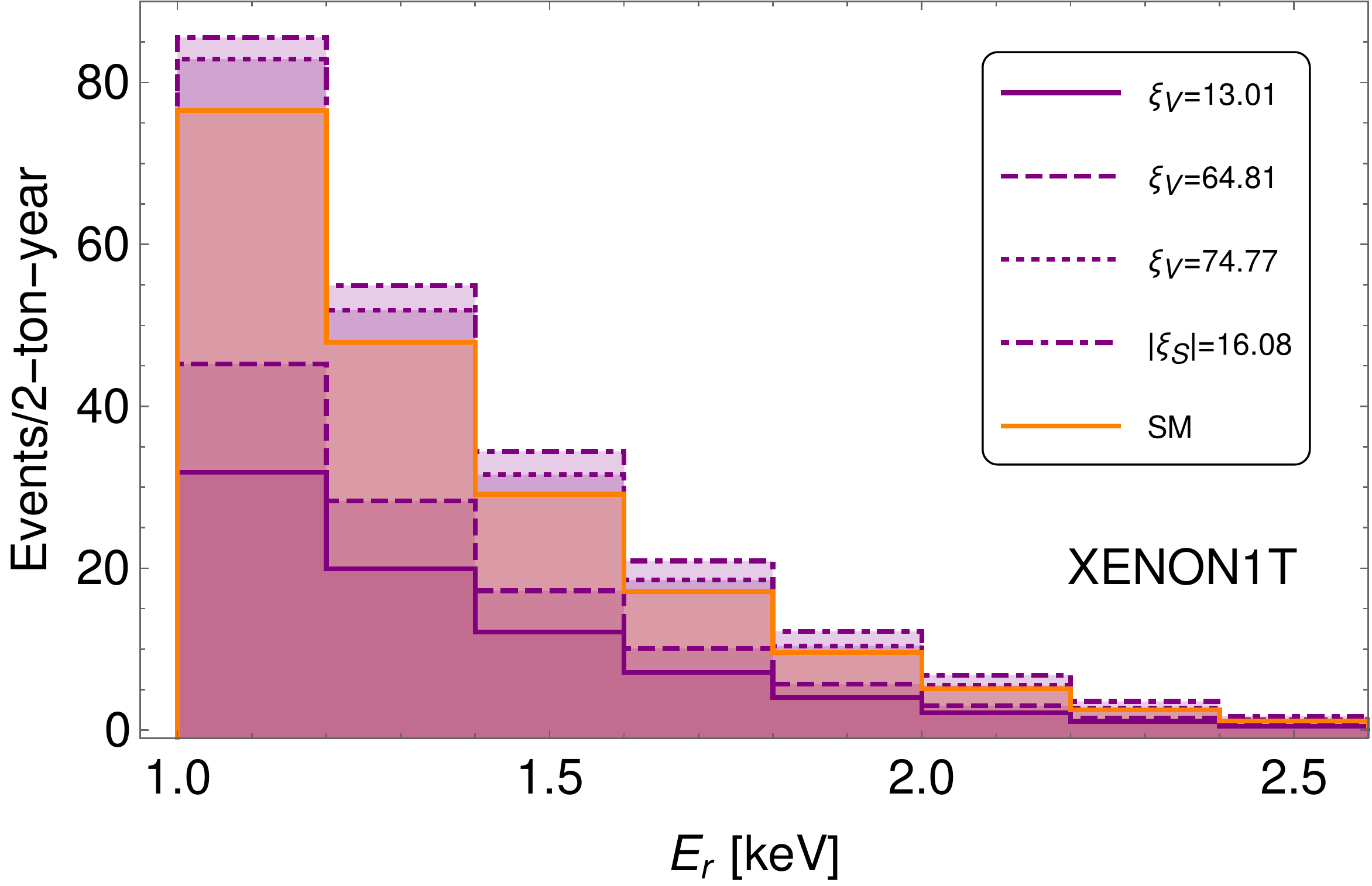}
  \caption{\textbf{Left graph}: Expected event rate spectrum for
    vector and scalar light mediators at XENON1T assuming a 2 ton-year
    exposure. Parameters are chosen to maximize the new physics event
    rate. Any other choice will therefore lead to events below these
    values. \textbf{Right graph}: Expected event rate spectrum for
    vector and scalar effective interactions. Results for LZ (10
    ton-year) can be directly derived by rescaling these event rate
    spectrum by a factor 5. For comparison the SM expectation is also
    shown (orange histogram). These results demonstrate that low
    recoil threshold measurements will provide sufficient statistics
    with which precise studies of new interactions can be done.}
  \label{fig:xenon1t}
\end{figure*}
In the case of light mediators silicon performs---in general---better
than argon and germanium. However, as shown in the bottom-left graph,
for vectors germanium is more sensitive to depletions than what
silicon and argon are. The question of xenon degeneracy breaking thus
depends to a large extent on the window where the parameters happen to
fall in. For germanium, in the range
$F_VC_V^N\gtrsim [0.5,3.0]\times 10^{-8}$ the signal is depleted below
the SM expectation with values as small as
$0.4\,N_\text{Event}^\text{SM}|_\text{SM}\simeq 167.1$.  For values
$F_VC_V^N\gtrsim 3.0\times 10^{-8}$ the signal gets enhanced, with
those enhancements leading to signals that can exceed the SM
prediction by three orders of magnitude in silicon. For light scalars,
values of the couplings at $10^{-8}$ lead to enhancements of order
$3$, as shown in the bottom-right graph. Above those values the signal
can exceed the SM expectation by more than a factor $10^2$, in silicon
as well.

After measurements in silicon are carried out they can be used as well
to distinguish vector from scalar interactions, at least in certain
regions of parameter space. Comparing the bottom-left and bottom-right
graphs in Fig.~\ref{fig:degeneracy} one can see that vectors produce
enhancements that scalars cannot reach. Thus measuring $R$ above
$\sim 200$ will provide an experimental prove that a new light vector
boson is at work. As we discussed in sec. \ref{sec:vec-scalar},
observation of depletions below the SM prediction will favor vector
interactions over scalar, and that applies in this case too. In the
region $R\subsetsim [1.1,200]$, one finds the same situation discussed
in sec. \ref{sec:vec-scalar}: A counting experiment alone cannot
distinguish vector and scalar signals. In that case information of the
recoil spectrum is required to gain information on the nature of the
new interaction.

We close this section by stressing that measurements of CE$\nu$NS in
xenon detectors matching the SM prediction not necessarily rule out
the presence of new physics. In that case our results encourage
measurements in any of the target materials we have considered, but
ideally in silicon.
\section{CE$\nu$NS and new interactions 
  in XENON1T and LZ}
\label{sec:new-interactions}
We now quantify the modifications that the interactions in
(\ref{eq:vector-NGI-light}) and (\ref{eq:scalar-NGI-light}) introduce
on the CE$\nu$NS event rate spectrum for on-going and future
Xenon-based experiments~\cite{Akerib:2018lyp,Aprile:2019bbb}. Since
for both cases we use a simplified acceptance function
$\mathcal{A}(E_r/\text{keV})=H(1-E_r/\text{keV})$, the differences
between the spectra arises only from exposure. For XENON1T we take
$\mathcal{E}_\text{Xe1T}=2\,\text{ton-year}$ while for LZ
$\mathcal{E}_\text{LZ}=10\,\text{ton-year}$. We therefore present only
results for XENON1T, results for LZ are obtained by a factor 5
rescaling.

With this detector specifications we then calculate the event rate
spectrum first for light mediator scenarios and then for effective
interactions. To do so we chose parameters that maximize
enhancements/depletions above/below the SM expectaction (see
Secs. \ref{sec:LM-VplusS} and \ref{sec:eff-int}). The results are
displayed in Fig.~\ref{fig:xenon1t}. For light mediators (left graph)
one can see that in both cases the number of CE$\nu$NS events can
readily exceed by far the SM expectation, provided thresholds are
pushed below 2 keV or so. For the parameter combinations that
maximizes the event rate we find
$N_\text{Events}^\text{Vec}\simeq 1643$ and
$N_\text{Events}^\text{Sca}\simeq 4340$, to be compared with the SM
expectation $N_\text{Events}^\text{SM}\simeq 189$. It is worth
stressing that lower thresholds, e.g $0.1$ keV as has been used for
the analyses in Sec. \ref{sec:vec-scalar}, tend to diminsh the
scalar-induced event rate compared to the vector. This can be easily
understood from the vector and scalar differential cross sections in
Eqs. (\ref{eq:x-sec-V}) and (\ref{eq:x-sec-S}). The scalar being
proportional to $E_r$ gets depleted at low thresholds, something that
does not happen with the vector. Thus, measuments at, say, 0.1 keV
somehow favor detection of vector interaction signals, on the contrary
at 1 keV detection of scalar signals is favored instead. The same
conclusion is found in the effective limits as well.

For effective interactions (right graph) results are rather different
as a consequence of the possible (maximum/minimum) values the
effective vector and scalar couplings can have, resulting in
deviations from the SM expectation not as pronounced as they are in
the light mediator case. Vector interactions can yield up to
$N_\text{Events}^\text{Vec}\simeq 205$, with that number diminished
down to $N_\text{Events}^\text{Vec}\simeq 112$ or
$N_\text{Events}^\text{Vec}\simeq 79$ depending on parameter
choice. Note that depletions below the SM expectation are found as
well in the vector light mediator case (see left graph in
Fig. \ref{fig:events}), so observation of event rates below those of
the SM will be a sufficient criteria to establish the vector nature of
the new interaction. For the scalar coupling case we find
$N_\text{Events}^\text{Sca}\simeq 221$, as expected always exceeding
the SM prediction. In summary, in both cases (vector and scalar) the
number of events exceeds (or goes below) the SM expectation with an
statistics that might be sufficient to identify the new interaction as
well as to eventually reconstruct its ``morphology''.
\section{Conclusions}
\label{sec:conclusions}%
In this paper we studied the sensitivity of multi-ton DM experiments
to distinguish vector and scalar new physics signals, through
measurements of CE$\nu$NS processes induced by $^8$B neutrinos. We
addressed the question of whether given a signal not matching the SM
expectation those measurements can identify the nature of the new
interaction (assuming the CE$\nu$NS signal can be differentiated from
a DM signal). We considered the case of light mediators as well as the
effective case. The former taken deep inside the light mediator window
$[1,10]\;$MeV, while the latter valid for $m_X\gtrsim 10^3\;$MeV.  We
used the recent COHERENT timing and energy data and neutrino mass
scales to apply constraints on both scalar and vector interactions for
light and heavy mediators before we investigate them at the DM
detectors.  Using a xenon-based DM detector we identified cases where
measurements of the event spectrum alone suffices to establish whether
the signal is due to vector or scalar couplings. For light mediators,
measurements yielding $R=N_\text{SM+Vec}/N_\text{SM}\gtrsim 84$ or
$R\lesssim 1$ will demonstrate the presence of a new light gauge
boson, while discarding a dominant light scalar contribution. In the
effective case, instead, measurements resulting in $R\gtrsim 1.05$ or
$R\lesssim 1$ will demonstrate that a ``heavy'' vector mediator is at
work.

We identified as well a region of degeneracy where a deviation above
the SM prediction can be accounted for by either the vector or the
scalar interaction. In the light mediator case we found that for
measurements resulting in $R\subsetsim (1,84]$, disentanglement of the
vector and scalar contributions cannot be done only by measurements of
the event spectrum. In the effective case we found that such region
exist as well for $R\subsetsim (1,1.05]$. We showed that such
vector-scalar signal degeneracy can be broken by combined measurements
of the event and recoil spectra. In the light mediator case
identification of vector interactions is possible because there is
always a dip in their recoil spectrum for recoil energies between
$0.1-3.5\;$keV, whereas for scalars this is never the case. In the
effective case, scalar interactions lead to recoil spectra that fall
more steeply than vector recoil spectra do. Combined measurements of
number of events and recoil spectra can then be used to identify the
origin of the new physics signal.

We considered as well the capability of multi-ton DM detectors to
measure the couplings of the new mediator to protons and neutrons, or
in other words to determine whether the new physics is or not isospin
conserving. For that aim we considered xenon, silicon, argon and
germanium detectors and focused on three extreme scenarios:
protophobia, neutrophobia and degeneracies. The latter defined as a
scenario in which the new physics exactly cancels in a particular
nuclide. We showed that in the effective case two independent
measurements of the event spectrum are sufficient to pin down the
value of the neutron-to-proton couplings ratio, and so to establish
whether the new physics conserves or breaks isospin. We demonstrated
that to establish the isospin nature of the new interaction, given a
first measurement in xenon, silicon is the most suited nuclide among
those we considered. We stressed that in contrast to measurements of
these type for DM, in this case statistics is---in principle---not an
issue and so implementation of such detector complementarity should be
feasible.

We studied the case of parameter space degeneracies assuming the new
physics signal exactly vanishes in xenon. We showed that degeneracy
breaking can be done in any of the detectors considered, but silicon
performs way better in both cases, light and effective limits. The
exception being only a particular region in parameter space for light
vector mediators, where depletions in germanium are more pronounced
than in argon and silicon. We pointed out that if measurements in
xenon match the SM expectation, efforts towards measuring CE$\nu$NS in
other ton-size detectors using silicon should be carried out to test
whether new physics is hidden in measurements involving xenon.

Finally, we found that the new vector and scalar interactions can be
investigated with $\sim 210-4000$ events (for parameters that maximize
the event rate) with a 2 ton-year experiment (e.g. XENON1T), or five
times those values for a 10 ton-year exposure (e.g. XENONnT and LZ),
if the nuclear recoil energy threshold is reduced down to 1
keV. Depletions below the SM expectation can be expected too. With
still sufficient statistics, $\sim 80-110$, they will demonstrate that
the origin of the new signal arises from a new vector force. Note that
the current thresholds $\sim$ 5 keV (XENON1T) and 4.4 keV (LZ) would
not apply any constraints on scalar and vector interactions in the
$^8$B solar neutrinos.  We also stress that the results presented here
apply as well to other CE$\nu$NS-related experiments such as CONNIE
\cite{Aguilar-Arevalo:2019jlr}, CONUS \cite{conus} and $\nu$-cleus
\cite{Strauss:2017cuu}. Interpretation of data from these experiments
in terms of new physics should take into account the results we have
presented in this paper.
\section*{Acknowledgments}
DAS is supported by the grant ``Unraveling new physics in the
high-intensity and high-energy frontiers'', Fondecyt No 1171136. He
would like to thank Nicolas Rojas for useful conversations. The work
of BD, SL and LS are supported in part by the DOE Grant
No. DE-SC0010813. DAS would like to thank the University of Texas A\&M
Mitchell Institute for Fundamental Physics for its hospitality.
\bibliography{references}
\end{document}